\documentclass[12pt]{article}
\usepackage{styles/mystyle} 

\begin{document}
\onehalfspacing
\title{Measuring the performance of investments in information security startups: An empirical analysis by cybersecurity sectors using Crunchbase data}

\date{\today}

\author[1,2]{Loïc Maréchal\thanks{Corresponding author: loic.marechal@unil.ch}}
\author[2]{Alain Mermoud\thanks{mermouda@ethz.ch}}
\author[3]{Dimitri Percia David\thanks{dimitri.perciadavid@hevs.ch}}
\author[1]{Mathias Humbert\thanks{mathias.humbert@unil.ch\newline\newline This document results from a research project funded by the Cyber-Defence Campus, armasuisse Science and Technology. We appreciate helpful comments from seminar participants at the Cyber Alp Retreat 2022 and WEIS 2023.}}

\affil[1]{\small Department of Information Systems, HEC Lausanne, University of Lausanne}
\affil[2]{\small Cyber-Defence Campus - armasuisse, Science and Technology}
\affil[3]{\small Institute of Entrepreneurship \& Management, University of Applied Sciences, HES-SO Valais-Wallis}

\maketitle

\newpage
\onehalfspacing
\begin{abstract}
    Early-stage firms play a significant role in driving innovation and creating new products and services, especially for cybersecurity. Therefore, evaluating their performance is crucial for investors and policymakers. This work presents a financial evaluation of early-stage firms' performance in 19 cybersecurity sectors using a private-equity dataset from 2010 to 2022 retrieved from Crunchbase. We observe firms, their primary and secondary activities, funding rounds, and pre and post-money valuations. We compare cybersecurity sectors regarding the amount raised over funding rounds and post-money valuations while inferring missing observations. We observe significant investor interest variations across categories, periods, and locations. In particular, we find the average capital raised (valuations) to range from USD 7.24 mln (USD 32.39 mln) for spam filtering to USD 45.46 mln (USD 447.22 mln) for the private cloud sector. Next, we assume a log process for returns computed from post-money valuations and estimate the expected returns, systematic and specific risks, and risk-adjusted returns of investments in early-stage firms belonging to cybersecurity sectors. Again, we observe substantial performance variations with annualized expected returns ranging from 9.72\% for privacy to 177.27\% for the blockchain sector. Finally, we show that overall, the cybersecurity industry performance is on par with previous results found in private equity. Our results shed light on the performance of cybersecurity investments and, thus, on investors' expectations about cybersecurity.
\end{abstract}


\thispagestyle{empty}
\clearpage

\onehalfspacing
\setcounter{footnote}{0}
\renewcommand{\thefootnote}{\arabic{footnote}}
\setcounter{page}{1}

\section{Introduction} \label{introduction}

\begin{quote}
If it were measured as a country, then cybercrime — which is predicted to inflict damages totaling \$6 trillion USD globally in 2021 — would be the world’s third-largest economy after the U.S. and China.\footnote{Cybersecurity Ventures, available at \href{https://cybersecurityventures.com/hackerpocalypse-cybercrime-report-2016/}{https://cybersecurityventures.com/hackerpocalypse-cybercrime}}
\end{quote}

In this 2020 article, \quotes{Cybersecurity Ventures} additionally expects the global costs of cyberattacks to increase by 15\% per year over the next five years, reaching USD 10.5 trillion annually by 2025. This represents the most outstanding transfer of wealth in history, more than the trades of all illegal drugs combined. To mitigate these costs, the information security business, especially cybersecurity, is expected to grow significantly. Indeed, these firms develop and implement new solutions to increase security in IT systems, protect virtual assets, custom sensitive information, or secure transactions and communications.

To better understand the maturity of the cybersecurity ecosystem and sustain its growth, it is critical to evaluate the financial performance of cybersecurity firms. First, their valuations and corresponding expected returns provide us with an almost direct metric for the potential of cybersecurity sectors. It also helps disentangle the different cybersecurity sectors regarding capital raised and valuations into which this capital translates, thus enhancing technology monitoring. Second, estimating financial performance parameters such as systematic financial risk (hereafter, risk) and risk-adjusted returns helps guide investors' decisions. This will eventually allow for an optimal capital allocation across target firms and should, in turn, improve the sector's financing and its ability to enhance the global level of cybersecurity.


In this research, we study the realized and estimated expected financial performance of private firms involved in cybersecurity sectors. We focus on the realized financial performance of 19 cybersecurity-related sectors identified in Crunchbase, a global database containing commercial and managerial data for private and public companies. In particular, we scrutinize their data about funding rounds, (post-money) firm valuations, and exits (IPOs or acquisitions). We restrict our sample over 2010--2022 for two reasons. First, Crunchbase is relatively recent, and many observations are missing before 2010. For instance, nearly 88\% of the recorded funding rounds occurred after 2010. Second, the magnitude of cyberattacks and the importance of the cybersecurity sectors have vastly increased since 2010. Crunchbase data is exhaustive regarding funding rounds. Still, many valuations and IPOs share price observations are missing. We use a machine learning approach to estimate them based on the dataset's numerous other highly correlated variables. We next follow Cochrane's (2005) approach to compute returns from financing round to exits, \textit{i.e.} IPOs or acquisitions~\cite{Cochrane2005}. We compute returns accounting for capital dilutions since several funding rounds before exits are common in the private equity sector.

We begin our analysis with descriptive statistics about the amount of capital raised and post-money valuations (PMVs) in the 19 sectors. The top three sectors based on the capital raised criterion are artificial intelligence, security, and machine learning, with more than USD 60 bln raised in each sector. Additionally, the private cloud sector dominates regarding average and median funding and post-money valuations, with up to USD 45.46 mln in average firm valuation. Regarding post-money valuations, we also find that artificial intelligence and machine learning dominate other sectors in total valuations. Yet, the private cloud sector also ranks first in average valuation with USD 447.22 mln, followed by the intrusion detection, QR codes, and fraud detection sectors. These sectors offer returns of up to USD 10 in share valuation for every USD invested. We confirm these results with a mean t-test analysis for pairwise comparisons of fundings and PMVs per sector. Regarding statistical significance, private clouds and intrusion detection sectors dominate others in valuations at the 1\% (10\%) level for up to seven (12) technologies. At the same time, E-signature and QR codes are the lowest-ranking sectors based on this criterion.
We then analyze our figures in terms of trends, setting up a quarterly grid and computing percent changes of average funding and valuations over the sample period. We find that the fraud detection sector had the most significant increase in funding amount (valuations) with over 283\% (586\%), followed by cloud security with 264\% (252\%) and privacy with 190\% (192\%). Instead, the lowest annualized average percent changes are associated with artificial intelligence, security, and machine learning. We sort capital raised and valuations by geographical locations and unsurprisingly find that most private equity investment in each sector is raised by US firms, except for Facial recognition (dominated by China with over 80\% of investment) and QR Codes (dominated by India with 98\% of investment). However, there are a few sectors where the US dominance is unclear, such as the penetration testing sector (with 41\% of funding going to Israeli firms) and the privacy sector (with Canadian firms capturing 26\% of investment). China also captures a significant market share in several sectors, such as E-signature, IoT, and biometrics.

We next assume a log return process and calculate returns to exit, accounting for capital dilution and quarterly returns for 12 cybersecurity sectors, estimating parameters and implied parameters. The $\alpha$ parameter, return in excess of the market risk premium, is positive for all technologies, while the $\beta$ parameter, indicating the systematic risk and cyclicality of the sectors, ranges between 0.51 and 5.51. We find sectors such as the artificial intelligence sector to be pro-cyclical, while firms in the more explicit cybersecurity sector have $\beta$ well below one, indicative of contra-cyclicality. Finally, we extract the implied expected arithmetic and log returns from the model parameters. We find the blockchain sector to have the highest expected annual arithmetic and log returns at 177.27\% and 105.42\%, respectively, consistent with the performance of cryptocurrencies over the sample period. The second-highest sector is artificial intelligence, with expected annualized arithmetic returns of 67.25\%, similar to Cochrane's (2005) findings for the information sector (79\%). Other sectors with high expected returns include machine learning, private cloud, and cloud security, whereas the lowest ranks include privacy (9.72\% p.a.) and biometrics (23.22\% p.a.).

To summarize, we find that in absolute terms, well-known sectors such as artificial intelligence or machine learning are dominant in terms of funding, whereas the private cloud dominates in terms of overall valuation. Nonetheless, in terms of average returns, the three aforementioned sectors remained slightly below the blockchain sector, although this performance is likely to be driven by investors' interest in cryptocurrencies.

Our contribution is twofold. First, in the economics of the information security field, we are the first to our knowledge to use financial metrics and benchmarking to assess the performance of cybersecurity sectors. Second, in the private equity literature, we are unaware of any research studying the industrial subsectors at such a granular level, even less in the specific, booming cybersecurity field. Overall, our results shed light on cybersecurity's financial performance as a whole but also support the view that they constitute a highly heterogeneous class of firms and that these sectors should be treated individually.

The remainder of this paper is organized as follows. Section 2 provides a literature review of the financial aspects of the economics of cybersecurity, its technical advances, and its peculiarity as a sector. Section 3 details the data and methods, Section 4 presents the results, and Section 5 concludes.

\section{Literature review and hypothesis development} \label{lit_review}
\subsection{Cybersecurity costs}

\subsubsection{Direct estimations}

Anderson et al. (2013) conduct a systematic study on the cost of cyber crimes~\cite{AndersonBartonBöhmeClaytonEetenLeviMooreSavage2013}. They disentangle direct, indirect, and defense costs and the types of cyber crimes. At the individual level, they find that typical online fraud directly costs citizens an amount of 100 USD per year. In contrast, they find that the indirect costs, spending for security to prevent these attacks, are an order of magnitude higher. They conclude that our societies are highly inefficient at preventing cyberattacks, given the vast externalities induced by these activities. Anderson et al. (2019) update this study to consider the fast evolutions that information systems have experienced, particularly the widespread uses of mobile phones and the changes in operating systems that followed, increased usage of cloud services, and social networks~\cite{AndersonBartonBöhmeClaytonGananGrassoLeviMooreVasek2019}. Even though payment frauds have doubled over the seven years separating the initial studies, their average costs for the citizen have fallen. They reestimate the costs of the various cyberattacks and, as Anderson et al. (2013) conclude that economic optimality would be in spending less on cyberattack prevention and more on response and law enforcement. Romanosky (2016) studies the structure and costs of cyberattacks and verifies whether incentives targeting firms exist to improve their security practices~\cite{Romanosky2016}. Using a sample of 12,000 cyber incidents, he examines the breach and litigation rate by industry and identifies the industries facing the greatest costs. He finds that the financial impacts are limited relative to the public concerns about these cyber incidents. He also estimates the typical cost of a cyber incident to be around USD 200,000 (similar to the firm's annual spending). Poufinas and Vordonis (2018) use actuarial methods to estimate cyberattack costs at the individual organization and aggregate levels~\cite{PoufinasVordonis2018}. Next, they estimate the equivalent premium a policy should require to insure against the discounted expected loss due to a cyberattack.

\subsubsection{Indirect estimations with stock price reactions}

The public equity market is one readily available proxy to estimate the costs of cyber attacks. While controlling for all confounding variables driving the future discounted cash flow expectations, the researcher can precisely estimate these costs using the cross-sectional and overall variations of the stock market around a cyber attack as investors revise their expectations. This method has been successfully employed to estimate the impact of FOMC announcements, policy rate revisions, or any other extreme event affecting the market. Moreover, from an econometric perspective, the models themselves have evolved to account for event-induced variance (\citealp{BoehmerMusumecciPoulsen1991}) cross-correlation (\citealp{KolariPynnonen2010}) in returns, or liquidity, among other counterfactuals (for an extensive literature review, examples, and methodologies, see, Campbell, Lo, and MacKinley (1997)~\cite{CampbellLoMacKinley1997}). Campbell et al. (2003) examine the cost of cybersecurity breaches on firms' market capitalization~\cite{CampbellGordonLoebZhou2003}. They use an event-study approach with a market model to build the benchmark. On an aggregate level, using all types of cybersecurity breaches, they do not find significant evidence for a negative effect of cyber attacks on firms' value. They repeat the study while disentangling the various cybersecurity breaches. In this setting, they find that information security breaches involving unauthorized access to confidential data yield a significant negative effect. In contrast, the breaches involving no confidential data do not yield significant abnormal negative returns. These results support the theoretical research of Kamiya et al. (2021), who derive a model assuming that a firm has optimal exposure to typical cyber risk~\cite{KamiyaKangKimMilidonisStulz2021}. In this setting, they find that a cyberattack should cause no financial and reputational harm to the firm. However, when extending the cyber risk to the context of the loss of personal information, the total cost of the attack, through indirect costs, becomes much larger. In addition, they find that an attack increases a firm's risk aversion. The model predicts that cyberattacks overall affect the stock price of the target firms as well as peer firms from the same industry.

Gordon et al. (2011) study the evolution of the incurred cybersecurity costs of breaches, repeating the event study approach of Campbell et al. (2003)~\cite{GordonLoebZhou2011}. They reemploy a market model as a counterfactual over a longer period. In contradiction with Campbell et al. (2003), they find that all breaches impact stock returns. They split the cybersecurity breaches into confidentiality, availability, and integrity events and find that availability breaches yield the most significant negative abnormal returns. Finally, splitting their sample into two sub-periods, before and after the 9/11/2001 attacks, they find that the significance of the negative abnormal returns observed around breaches decreases in the post-period. They explain these results by improving firms' disaster recovery, and the public awareness of the real costs for the expected future cash flows \textit{vs.} what media reports. Tosun (2021) studies cumulative and buy-and-hold abnormal returns in another event study setting~\cite{Tosun2021}. He supports the previous evidence of significant abnormal returns and other understudied features. In particular, cyberattacks are associated with increased trading volume due to selling pressure and improvements in liquidity, but only when the firm is targeted for the first time by a cyber event. Finally, he finds that cyberattacks affect the firm's governance for up to five years after the announcement of the cyberattacks. Lending et al. (2018) study the relation between cyberattacks and corporate and social responsibility (CSR)~\cite{LendingMinnickSchorno2018}. They show that well-ranked firms in the latter criterion are less likely to be harmed. They also estimate the long-term abnormal returns on a firm's share to be around -3.5\% per year. Finally, they find a positive externality in cyberattacks, with companies more likely to improve their CSR following a cyberattack (also see Spanos and Angelis; 2016 for a literature review of the impact of security events on the stock market and Johnson et al.; 2013)~\cite{SpanosAngelis2016,JohnsonKangLawson2017}.

\subsubsection{Indirect estimations with disclosures}

Gordon et al. (2010) study the impact of voluntary disclosures on public firms regarding information cybersecurity in SEC filings~\cite{GordonLoebSohail2010}. As for many disclosures, information security is purely voluntary. A significant strand of accounting research relates empirically and theoretically voluntary disclosures to increased market valuations through the liquidity channel. They use a panel of 1,641 firms disclosing information security \textit{vs.} 19,266 firms that do not. They find strong evidence that information security disclosures positively relate to higher market value. In addition, using the bid-ask spread size as a proxy for liquidity, they find that the market valuation increase is due to this liquidity channel. Whereas the economic model of Laube and Böhme (2016) foresees few incentives for firms to disclose, the market valuation reason could mitigate this view~\cite{LaubeBohme2016}. Amir et al. (2018) use data on firms' voluntary disclosures around cyberattacks and withheld information that is later disclosed~\cite{AmirLeviLivne2018}. They identify that managers' incentives to retain information make them under-report and find that cyberattacks generate a 3.6\% decline in market capitalization over the following month. When the cyberattack is disclosed, firms undergo a further decline of 0.7\%. Finally, they support the fact that managers disclose information only as investors estimate the probability of attack to be above 40\%. Hilary et al. (2016) study misalignments between regulators' incentives for corporate disclosure related to cybersecurity risk and the reality~\cite{HilarySegalZhang2016W}. They find that cyber risk disclosures of public firms are rare and that breaches have a limited impact on stock price effects. However, their qualitative study is limited to only five significant cases, limiting support for their claims. First, given the small size of their sample, and second, because the limitations to major cases are likely to induce a selection bias. Finally, Florackis et al. (2022) use textual analysis on cybersecurity risk disclosures extracted from 10-K SEC filings~\cite{FlorackisLoucaMichaelyWeber2023}. Identifying that the amount of disclosures related to cybersecurity predicts future cyberattacks, they rank firms on this criterion and find that firms with high exposure to this factor exhibit significantly higher expected returns than those without. Moreover, they find that this cybersecurity risk premium is not subsumed by the traditional financial risk factors and is orthogonal to them. Thus, cybersecurity risk is priced in the cross-section of firms, with applications for policymakers, risk, and portfolio managers.

\subsubsection{Cost metrics}

Wang et al. (2008) adopt a value-at-risk approach, a standard metric in the financial risk management literature and industry, to overcome the problems of traditional metrics, such as annual loss expectancy, which do not appropriately fit the profile of the rare, yet extremely costly nature of cyberattacks~\cite{WangChaudhuryRao2008}. They estimate the value-at-risk using extreme value theory and calibrate their model on daily data from large firms. They refine the models using information from interviews with security managers and estimate firms' potential daily loss. This would, in turn, allows decision-makers to optimize their cybersecurity investment decisions based on this metric (also see, Bouveret; 2018 for a discussion about the value-at-risk framework adoption in the financial sector)~\cite{Bouveret2018W}. Along the same line, Bodin et al. (2008) use the \quotes{perceived composite risk} to overcome the issues with traditional metrics such as expected loss from a breach or the standard deviation of the expected loss~\cite{BodinGordonLoeb2008}. Finally, Ruan (2017) coins the term \quotes{cybernomics}, which would gather all the aforementioned risk measures (among others) and economic frameworks (also see Fielder et al.; 2018))~\cite{Ruan2017,FielderKonigPanaousisSchauerRass2018}.

\subsection{Cybersecurity investments}

\subsubsection{Investment in cybersecurity technologies}

As for all security economics, modeling optimal decisions in the cybersecurity context is an actuarial problem. Gordon and Loeb (2002; hereafter, GL) derive a cybersecurity investment model to determine the optimal amount for protection~\cite{GordonLoeb2002}. Endogenizing information vulnerability to a cyber attack and the potential loss if the breach occurs, they find that the most vulnerable information may not be worth protecting as the costs could be too high. With their model, they advocate investing in protecting midrange vulnerabilities to reach optimality. They additionally find an aggregate optimal investment level in cybersecurity of 37\% of the expected loss due to a cyber attack. This latter result is crucial as much of the empirical research finds support for under-investments in cybersecurity, and this is for all kinds of organizations. Gordon et al. (2015a)extend the GL model to potential externalities of cybersecurity breaches~\cite{GordonLoebLucyshynZhou2015a}. They consider, in particular, attacks that would affect not only individual organizations but also an entire critical infrastructure industry. Under this setting and the change in social welfare induced by such attacks, they revise the original GL model and find that the firm's optimal social investment in cybersecurity increases by, at most, 37\% of the expected externality loss. Gordon et al. (2015b) uncover a positive relationship between firms' information sharing and their proactivity towards cybersecurity investments~\cite{GordonLoebLucyshynZhou2015b}. They show how information decreases the deferring of cybersecurity investments. They consider the cybersecurity investment a real option, a call option exercised through future loss avoidance. Put differently, the option value to defer cybersecurity investments increases as the uncertainty increases. In this setting, they find that the uncertainty reduction through information sharing decreases the value of the option deferment, thereby moving forward the investment decision. Farrow and Szanton (2016) extend GL, including changes in the probability of attacks, event-induced variance (simultaneous effects on probability and loss), diversification of attack, and shared non-information defenses~\cite{FarrowSzanton2016}. Willemson (2006) studies whether an upper limit exists for the level of optimal security investments in the GL context~\cite{Willemson2006}. He disproves the GL conjecture, which intuitively sets the level at around $\nicefrac{1}{e} = 36.8\%$, and studying several examples and relaxing the original hypotheses, this level can vary between 50\% and 100\%. Instead, Baryshnikov (2012) analyzes a major criticism addressed to GL regarding whether their $\nicefrac{1}{e}$ rule holds after the model is generalized~\cite{Baryshnikov2012}. In this context, he proves that the rule does hold in full generalization (see also, Hausken (2006)).\footnote{Building upon the GL model, a vast literature in economic modeling of cybersecurity has arisen. Other research includes Gordon et al. (2016), who discuss the GL Model and its applications in a practical setting, Gordon and Loeb (2006), and Anderson and Moore (2006), who summarize advances in the economic aspects of information security, or Rue and Pfleeger (2009) who compare five investment models in information security~\cite{Hausken2006,GordonLoebZhou2016,GordonLoeb2006,AndersonMoore2006,RuePfleeger2009}}

\subsubsection{Economic incentives for cybersecurity investment}

Gordon et al. (2015c) derive an economic model incorporating government incentives to mitigate private firms' underinvestment in cybersecurity-related activities. They show that the efficiency of these incentives depends on the firm's usage of an optimal mix of cybersecurity inputs and their willingness to increase their investments in cybersecurity~\cite{GordonLoebLucyshynZhou2015c}. Lelarge (2012) examines the issue of incentivizing agents in a large network to invest in cybersecurity~\cite{Lelarge2012}. Starting from a single agent, he derives the optimal invested amount in protection while considering potential losses and extending it to a network of agents. He finds that security investments are generally socially inefficient due to network externalities and that aligning incentives typically leads to a coordination problem and, thus, a high price of synchronicity. Wang (2019) models the optimal investment in cybersecurity and insurance based for a firm and demonstrates how private sector efforts reduce the aggregate cyber loss for society~\cite{Wang2019}. In this setting, the incentives come from positive externalities at the global level. He also disentangles the various benefits of the cybersecurity efforts given firms' sizes. He finds that small and medium-sized firms benefit the most from cyber insurance (also see Gordon et al. (2003) for a discussion on a framework in cyber insurance)~\cite{GordonLoebSohail2003}. Extensions of the GL model also integrate the cost-benefit analysis into the existing framework for cybersecurity risk management. For instance, Gordon et al. (2020) use the GL model to complement the NIST cybersecurity framework~\cite{GordonLoebZhou2020}. In this aspect, the NIST framework, which accounts for the peculiarities of organizations, is a relevant framework, albeit lacking a cost-benefit part. Combining the two approaches, they show that a system based on the GL model allows one to select the NIST Implementation Tier level for each organization optimally and to identify the optimal timing to switch from one NIST level to the next.

\subsubsection{Empirical evidence of investment}

Few studies analyze the actual level of cybersecurity investments. One exception is Gordon et al. (2018), which circumvents the analysis to the private sector~\cite{GordonLoebLucyshynZhou2018}. They tackle the outcome of a survey assigned to private firms to assess the relative importance of internal control systems in their financial reports. One component studied is cybersecurity, and it is thus used as a proxy for the firm's willingness to invest in cybersecurity activities. They additionally show that a firm's concern towards cybersecurity risk and its perception of cybersecurity investment as leading to a competitive advantage are positively related to the level of cybersecurity investments. Nonetheless, they find that private sector firms underinvest in cybersecurity and explore the causes. Moore et al. (2016) assign a survey in the form of semi-structured interviews conducted with 40 information security executives and managers of firms from the healthcare, financial, retail, and governmental sectors~\cite{MooreDynesChang2016}. They find that 81\% of the interviewees assess that company management supports cybersecurity efforts and that this trend is increasing (85\%). Moreover, 88\% of the participants report that their cybersecurity budget has increased. These results contradict the empirical evidence of under-investment and lack of willingness at the aggregate level. However, one cannot discard that the self-perception of the interviewees and psychological biases may drive these good results. Tanaka et al. (2005) empirically study the link between vulnerability and the level of investment in information security~\cite{TanakaMatsuuraSudoh2005}. To do so, they use data recorded from the \quotes{Census of e-Local Government} in Japan, which covers 3,241 municipal governments. This dataset provides them with a more direct proxy than the traditional IT factors generally used (hardware, software, outsourcing, and workers in IT) as they directly observe the investment policy and attribute it to a dummy variable. They find these policies are significantly and positively related to the (log) level of information security investment. Romanosky (2016) estimates the typical firm's annual cybersecurity spending is around USD 200,000~\cite{Romanosky2016}. In contrast, he finds this amount to represent only 0.4\% of the firm's typical annual revenue.


\subsection{Financial methods for venture capital}

\subsubsection{Estimation challenges in venture capital}

Studies in private equity markets face two problems. First, as opposed to public companies, private firms are not legally required to disclose their financial statements. For instance, in the US, this is done through the security and exchange commission (SEC) forms 10-Q (quarterly) or 10-K (annually). In addition, public firms must communicate through an annual report to their shareholders. While none of this is mandatory for private firms, estimating financial performance is difficult without insider information. Second, private firms have no public shares to trade, and we do not observe any quasi-continuous price processes or market capitalization. Thus, almost none of the estimations employed on the public equity market are readily usable. Nonetheless, private firms issue shares to investors in various forms, with different optionality clauses. Private equity analysts rely on insider or private information to evaluate projects. The common practice is to value a company around financing events.\footnote{A financing event is any event during which the firm receives equity, issues debt or receives grants.} This valuation is called pre- (post-)money valuation when done before (after) the financing event. A private equity analyst can obtain the firm valuation by multiplying the per-share price of the most recent event by the fully diluted number of common shares. This calculation, however, does not account for the optionality of the investment contract and assumes that all shares have the same value, regardless of their type (common, preferred, or convertibles notes). See, \textit{e.g.}, Gornall et al. (2020)~\cite{GornallStrebulaev2020}.

\subsubsection{Venture capital valuation}

Whereas the quasi-continuous processes of prices observed on the public equity market only require linear regressions to estimate the usual financial performance metrics (expected returns, systematic and idiosyncratic risk) of individual firms, this estimation is more cumbersome in private equity. Cochrane (2005) uses a maximum likelihood estimation method to obtain these values~\cite{Cochrane2005}. He analyzes the whole market and particular sectors such as healthcare and biotech, tech companies, and retail services. He finds a mean arithmetic return of 59\%, an alpha of 32\%, a beta of 1.9, and a volatility of 86\% (corresponding to a 4.7\% daily volatility). Because of the returns distribution, which is heavily positively skewed, he fits a logarithmic model. Given those observations for successful firms are likely to be over-represented in its sample, he adapts his model to include a selection bias parameter directly in the likelihood function. Ewens (2009) updates the methodology but focuses on round-to-round returns~\cite{Ewens2009}. He splits the logarithmic model into a three-regime mixture model (failure, medium returns, and \quotes{home runs}) and a separate holding period model to simulate the time between two financing events. He obtains an alpha of 27\% and a beta of 2.4. He finds that 60\% of all private equity investments have negative returns and high idiosyncratic volatility. Korteweg and Nagel (2016) extend the public market equivalent (PME) method to assess the performance of private equity funds~\cite{KortewegNagel2016}. They estimate monthly arithmetic alphas of 3.5\%, in line with previous results. They argue that their method delivers similar results more simply with no distributional assumptions. Moskowitz et al. (2002) obtain more nuanced results and conclude that returns on private equity are not higher than public equity~\cite{MoskowitzVissing-Jørgensen2002}. By finding a high idiosyncratic risk of single private firms, they conclude that the aggregate return overestimates the average returns to investors.

In a second strand of research, Alexon and Martinovic (2015) and Franzoni et al. (2012) estimate abnormal returns and risk factor loadings with standard regressions by using either internal rates of return (IRRs) or modified internal rates of return (MIRRs)~\cite{AxelsonMartinovic2015,FranzoniNowakPhalippou2012}. Driessen et al. (2012) and Ang et al. (2018) present an approach that extends the (IRR) calculation to a dynamic setting in which they solve for the abnormal returns and risk exposures using the Generalized Method of Moments (GMM). This approach requires only a cross-section of observable investment cash flows~\cite{DriessenLinPhalippou2012,AngChenGoetzmannPhalippou2018}.

The third line of research attempts to identify successful features of exits or features explaining venture valuations. Cumming and Dai (2011) identify a convex relationship between fund size and firm valuations and a concave one between fund size and a target company's performance~\cite{CummingDai2011}. Similarly, Cumming and Dai (2010) examine local bias in venture capital investments~\cite{CummingDai2010}. Based on a sample of US investments from 1980 to June 2009, they find that funds with broader networks exhibit less regional bias. More importantly, they relate geographical distance to the performance of VC investments and, thus, to the valuation of the target firm. Finally, Engel and Keilbach (2007) study the relationship between patents and early-stage firms~\cite{EngelKeilbach2007}. They find that venture-funded firms have more patent applications than those in the control group before investment. However, this relationship does not hold after investment. This suggests that venture capitalists focus on the commercialization of existing innovations.

The literature is sparse on the topic, and there is no consensus on methodology and results. Nonetheless, previous works estimate similar values. Cochrane's (2005) method stands out by its simplicity and has been successfully used in later research. It also focuses on returns from rounds to IPO, which is not the case for Korteweg and Sorensen (2010)~\cite{KortewegSorensen2010}.\footnote{Another strand of the literature not covered here concerns the construction of indices and benchmarks related to private equity performance. See, \textit{e.g.}, Peng (2001), Hwang et al. (2005), Schmidt (2006), Cummung et al. (2013), and McKenzie et al. (2012)~\cite{Peng2001W,HwangQuigleyWoodward2005,Schmidt2006,CummingHaßSchweizer2013,McKenzieSatchellWongwachara2012}.}

\subsubsection{Cybersecurity sector performance}


We are only aware of one study that focuses on the performance of the cybersecurity sector. Mezzetti et al. (2022) use a bipartite graph approach to link early-stage cybersecurity firms to technologies and ranks them based on features such as investors' type or geographic distance between investors and firms~\cite{MezzettiMarechalPercia-DavidLacubeGillardTsesmelisMaillartMermoud2022W}. However, the ranking is relative and depends on investors' preferences. To fill the gap in the absolute ranking of cybersecurity sectors and test their financial performance, we write our null hypotheses as follows,\\\\

\noindent
H1 Cybersecurity sectors are homogeneous in terms of:\\
H1a capital raised.\\
H1b firms' valuations.\\
H2 There is no difference in financial performance (risk-adjusted returns, systematic risk, and expected returns) across cybersecurity sectors.\\
H3 The financial performance of cybersecurity does not differ from that of broad private equity.\\

To test these hypotheses and disentangle the performance of the cybersecurity sub-sectors, we borrow from the approach of Cochrane (2005)~\cite{Cochrane2005}. We choose to estimate our parameters in private equity for three reasons. First, we can observe a larger cross-section of firms, w.r.t. the public equity. Second, the likelihood of cybersecurity firms having a single core business in the target sub-sector is much higher for small and middle firms that are typically private. Last, cybersecurity businesses are generally smaller, and few firms are listed, which would further limit the number of observations. As anecdotal evidence, the S\&P 500 does not include any component being a pure cybersecurity company (despite having four out of five top components being tech companies).


\section{Data and methodology} \label{data_methodology}
\subsection{Data}\label{section:cb}
\subsubsection{Crunchbase}

Crunchbase is a global commercial financial and managerial data vendor for private and public companies. Created in 2007 by TechCrunch, it has been maintained by Crunchbase Inc. since 2015~\cite{Crunchbase2022}. This database is used by academics, NGOs, and industry practitioners~\cite{Besten2021,DalleBestenMenon2017W}. Crunchbase collects data daily by combining crowd-sourcing, NLP-based newswire analyses, and in-house processing. It also complements it with data from third-party providers. The dataset is organized into several entities, including those relevant to this study. First, \quotes{organizations} is the set reporting administrative information on private and public companies, funds, or institutions. It includes business information, contact details, location, number of employees, and sector of activity. Second, \quotes{people} is the set containing information about individuals involved with a firm. It includes age, CV, degrees, or gender. Third, \quotes{funding rounds} is the set including funded companies, investors, round types (seed, series A/B/C, \ldots, or debt issuing), and amount of money raised. Fourth, \quotes{exits} reports the firm's acquisitions and their types (LBOs, management buyouts, or mergers) as well as IPOs-related information (new listing or delistings, share prices, market capitalization, and exchange venue). Crunchbase includes additional business information irrelevant to our research and thus not discussed here. 

One caveat of the dataset is that many observations regarding funding rounds are missing. In particular, post-money valuations (PMVs) and IPO share prices, which are central to our study, are often unavailable. Moreover, the multiple sources of information for Crunchbase data may induce heterogeneity, and the observations' quality will likely vary across country, industry, or period. For instance, US companies are over-represented compared to other nations. Finally, the observations were almost inexistent before the 2000s and sparse before the 2010s. However, we assume this latter issue to mildly affect our study since our goal is to estimate the current financial performance of cybersecurity sectors. Moreover, Crunchbase has a stronger focus on the technology industry, which is an advantage for this research.

We download data until May 2022; nearly 88\% of the recorded funding rounds occurred after 2010, and the trend is upward. This is due to the data availability and the fact that the VC market has vastly increased after the 2008 global financial crisis.

\subsubsection{Market data}\label{section:md}

We fit the model on one public equity benchmark and one risk-free asset. We use two sources: Yahoo Finance \footnote{Yahoo Finance, \url{https://finance.yahoo.com/lookup/}} and the Federal Reserve Bank of St. Louis (FRED). From their APIs, we collect the returns on the S\&P 500 index. For the risk-free asset, we use the 3-month T-Bill rate \footnote{Board of Governors of the Federal Reserve System (US), 3-Month Treasury Bill Secondary Market Rate, Discount Basis [TB3MS], retrieved from FRED, Federal Reserve Bank of St. Louis; \url{https://fred.stlouisfed.org/series/TB3MS}}, which helps for our analysis since we use a time grid of three months to fit the model. Given the low frequency of observations, it is also the standard in the existing private equity research.

\subsection{Methodology}

\subsubsection{Missing data interpolation}

\begin{figure}
  \centering
    \includegraphics[scale=0.7]{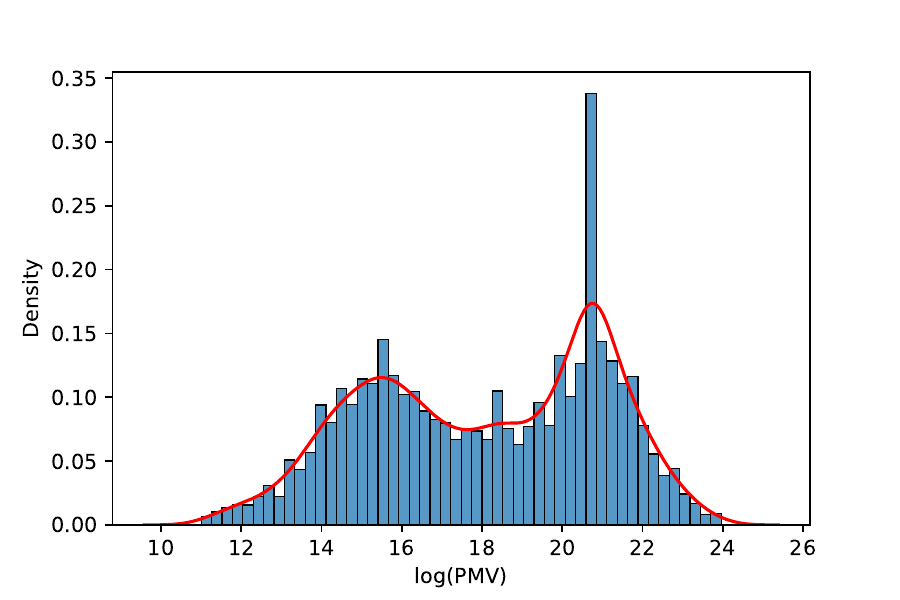}
    \caption{PMVs distributions from Crunchbase firms of all categories}
    \footnotesize{This Figure depicts the distribution of the logarithm of all PMVs of firms of all industries available in Crunchbase. The bars depict the discrete density; the red line is the kernel distribution. The period is 2010--2022.}
    \label{cb:log_pmv}
\end{figure}

To circumvent the problem of missing PMVs, and to a lesser extent, missing funding rounds amount, we interpolate the missing data leveraging the other features of the database using an ML regression approach. To leverage all the information in the database, we use the entire set, including firms that do not belong to the cybersecurity sub-sectors, to estimate the missing PMVs. We are not interested in the precise value of the firm after a financing event but rather in an unbiased estimate of the order of magnitude of the business value. We plot the distribution of the PMVs reported by Crunchbase for all firms in Figure \ref{cb:log_pmv}. Note that PMVs follow a bi-modal distribution, centered roughly around the 2-3 million value ($e^{15} \approx 3e6$) and the billion value ($e^{21} \approx 1e9$). This is likely due to the round number bias studied in \textit{e.g.} Herve and Schwienbacher (2018). Financial analysts tend to stick to round numbers because of the lack of firm information and the high valuation risks involved~\cite{HerveSchwienbacher2018}.

\begin{table}[ht]
    \centering
    \begin{tabular}{cl}
        \textbf{Symbol} & \textbf{Description}                                      \\ \cmidrule(lr){1-1}  \cmidrule(lr){2-2}
        \textbf{T}              & Date of the round (in days, relative to 1/1/1926) \\ 
        $\mathbf{\Delta T}$     & Number of days since the last financing event         \\ 
        \textbf{M}              & Amount of money raised (\$)                       \\ 
        $\mathbf{\Delta M}$     & Difference of money raised since the last round (\$)  \\ 
        \textbf{N}              & Number of investors for the current round         \\ 
        \textbf{R}              & Lead investor rank for the current round          \\ 
        \textbf{S}              & Industry sector (categorical)                     \\ 
        \textbf{G}              & Geographical position (categorical)               \\ 
    \end{tabular}
    \caption{Features used for the inference with PMVs regression}
    \footnotesize{This Table lists the features selected from the Crunchbase dataset to model the missing observations of PMVs in certain funding rounds.}
    \label{tab:pmv_features}
\end{table}

We report the features in Table \ref{tab:pmv_features} and capitalize on the high correlation between funding round size and PMVs, and between funding round frequency and PMVs, respectively. Also see Alexy et al. (2012)~\cite{AlexyBlockSandnerTerwal2012}.\footnote{The particular implementation used in this project is \quotes{AutoSklearn}~\cite{FeurerEtAl2021}.}

\subsubsection{Taxonomy and classification}

One challenge of our research is to map the Crunchbase taxonomy above to existing taxonomies. Whereas taxonomies and classifications of cyber events are numerous (see,\textit{e.g.}, Agrafiotis et al. (2018) for a taxonomy of cyberattacks and Shameli et al. (2016) for a taxonomy of information security risk assessment (ISRA), and NIST, those for the corresponding cybersecurity technologies mitigating these threats is not readily available~\cite{AgrafiotisNurseGoldsmithCreeseUpton2018,Shameli-SendiAghababaei-BarzegarCheriet2016,NIST2018}.\footnote{Despite the more extensive availability of cyberattacks taxonomies, Ruan (2017) advocates for a consistent taxonomy of cyber incidents~\cite{Ruan2017}.}

We identify two classifications that help us select the Crunchbase tags to allocate each firm's observations, such as funding rounds and PMVs, to the cybersecurity sub-sectors. One comes from the European Union Agency for Cybersecurity (ENISA), more specifically, from the ENISA cybersecurity market analysis framework (ECSMAF).\footnote{\url{https://www.enisa.europa.eu/publications/enisa-cybersecurity-market-analysis-framework-ecsmaf}} The second comes from Hwang et al. (2022). We present a reproduction of their classifications in Appendix,  Table \ref{tab:ENISA} and Table \ref{tab:Hwang}, respectively~\cite{HwangShinKim2022}. Based on these classifications, we identify the Crunchbase tags in the database that we assume to be related to cybersecurity sectors. The identified tags are the following,\\

\textit{Artificial Intelligence, Biometrics, Blockchain, Cloud Security, Cyber Security, E-Signature, Facial Recognition, Fraud Detection, Internet of Things, Intrusion Detection, Machine Learning, Network Security, Penetration Testing, Privacy, Private Cloud, QR Codes, Quantum Computing, Security, Spam Filtering}\\

We know that Artificial intelligence and Machine learning cover fields that go well beyond the cybersecurity sector. However, we choose to include them for two reasons. First, as for other fields, they take increasing importance in the cybersecurity sector, for instance, internet traffic analyses or fraud detection. Second, this inclusion will not bias our results regarding other sectors because all studies are done per sector.

\subsubsection{Financial method}

To compute returns from individual funding rounds to exits (acquisitions or IPOs), the single case considered in this study, we take capital dilution into account. First, we calculate the equity value at the exit for an investor entered at a funding round $i$:
\begin{equation*}
    x_i = \underbrace{\frac{m_i}{v_i}}_{\substack{\text{initial stake} \\ \text{of investors i}}} \times \underbrace{\frac{v_i - m_{i+1}}{v_{i+1}}}_{\substack{\text{proportion of old} \\ \text{equity at round } i+1}} \times \ldots \times \underbrace{\frac{v_{n-1} - m_n}{v_n}}_{\substack{\text{proportion of old} \\ \text{equity at round } n}}
\end{equation*}
Where $m_i, m_{i+1}, \ldots, m_n$ are the amount raised from investors at each round, $v_i, v_{i+1}, \ldots, v_n$ is the equity value at each round, and $x_i$ is the percentage of equity owned by investors on exit. The return for these investors is,
\begin{equation*}
    R_i = \frac{\overbrace{v_n \times x_i}^{\substack{\text{value owned}\\\text{at exit}}} - m_i}{m_i}
\end{equation*}
Next, we use a similar approach to Cochrane (2005)~\cite{Cochrane2005}. We do not rely on his maximum likelihood approach as fewer observed returns with disaggregated sectors would make the optimizer convergence infeasible. Instead, we scale returns at the exit date by the time difference in days between the initial funding round and observed or inferred PMV and the exit time. We obtain implicit daily arithmetic returns that we convert into quarterly returns. We subsequently average these returns in each cybersecurity sector to obtain our time series of returns. Finally, the particularly skewed distribution of private equity returns imposes the use of a log model,
\begin{equation}\label{eq_1}
    \begin{cases}
        \rm{d}\ln V = (r^f + \gamma)\rm{d}t + \delta(\rm{d}\ln V^m -r^f\rm{d}t) + \sigma \rm{d}B\\
        \rm{d}\ln V^m = \mu_m \rm{d}t + \sigma_m \rm{d}B^m
    \end{cases}
\end{equation}
Where $\rm{d} B$ is a standard Brownian motion, $V$ is the value of the asset (firm capital), $r^f$ is the risk-free rate, $\gamma$ the intercept, $\delta$ the slope, $V^m$ is the value of the market and $\sigma$ is the volatility of the value process. We follow Cochrane (2005) and assume $[B, B^m]_t = 0$~\cite{Cochrane2005}. In discrete time (for a time step $\Delta t = 1$), the model is,
\begin{equation}\label{eq_2}
    \ln\left(\frac{V_{t+1}}{V_t}\right) = \ln R^f_{t+1} + \gamma + \delta(\ln R^m_{t+1} - \ln R^f_{t+1}) + \epsilon_{t+1}
\end{equation}
Where $\epsilon_{t+1} \sim \mathcal{N}(0, \sigma^2\Delta t)$,  $R^f_{t+1} = 1 + r^f_{t+1}$, and $R^m_{t+1} = 1 + \frac{V^m_{t+1} - V^m_t}{V^m_t}$
Thus, the value of the firm $V_{t+1}$ follows a log-normal distribution with parameters:
\begin{align*}
    &\mu_{t+1} = \ln R^f_{t+1} + \gamma + \delta(\ln R^m_{t+1} - \ln R^f_{t+1})\\
    &\sigma_{t+1}^2 = \sigma^2
\end{align*}

\begin{equation}\label{eq_3}
    \mathbb{E}[\ln r] = \gamma + \mu_{\ln r_f} + \delta(\mu_{\ln r_M} - \mu_{\ln r_f})
\end{equation}

\begin{equation}\label{eq_4}
    \mathbb{V}[\ln r] = \delta^2\sigma_{\ln r_M}^2 + \sigma^2
\end{equation}

Since we use a quarterly frequency, we multiply by four to annualize results and by 100 to get percentages. To get the results for the arithmetic returns, we take the expectation and variance of a log-normal variable, with $R = r + 1$ and $\mu = \mathbb{E}[\ln R]$:

\begin{equation}\label{eq_5}
    \mathbb{E}[R] = \exp\left(\mu + \frac{1}{2}\sigma^2\right) - 1
\end{equation}

\begin{equation}\label{eq_6}
    \mathbb{V}[R] = (\exp(\sigma^2) - 1)\exp(2\mu + \sigma^2) = (\exp(\sigma^2) - 1)(\mathbb{E}[R] + 1)^2 
\end{equation}

\section{Results} \label{results} 
\subsection{Descriptive statistics}

We start by reporting descriptive statistics of our data, funding amounts, and PMVs. Table \ref{tab_descr} reports the number of firms, the number of funding rounds targeting the firms, and the average, median, total values, and standard deviations for each cybersecurity sector's funding amount and PMVs. The top three are the artificial intelligence, security, and machine learning sectors, with USD 124.3, 67.5, and 67.3 bln raised over 2010--2022, respectively. We know these three sectors might include more firms than those using these technologies for cybersecurity reasons, particularly artificial intelligence and machine learning. Regarding security, however, the fact that Crunchbase is heavily tilted towards tech companies, together with a manual sampling of the firms to check their actual businesses, makes us confident that a large part of them relates to cybersecurity or the more comprehensive information security. They also dominate the ranking in funding rounds events, with up to 6,339 rounds recorded for the Artificial intelligence sector. The following category in this ranking is the more specific cybersecurity category, with around USD 40 bln raised over the last decade, followed by blockchain, with over USD 27 bln. Once again, for this latter category, we are aware of the biases that may be induced by cryptocurrencies and NFTs companies that do not relate to the pure cybersecurity sector. However, the technology itself is tightly connected to it, and Crunchbase has other denominations for pure financial-related projects in the space, such as \quotes{Cryptocurrency}, \quotes{Bitcoin}, or \quotes{Ether}. The private cloud sector vastly dominates other sectors regarding average (USD 45.46 mln) and median funding (USD 11 mln). Following the average funding criterion, it is followed by QR codes (USD 37.76 mln), facial recognition (USD 31.48 mln), and cloud security (USD 25.23 mln). Instead, spam filtering, biometrics, and IoT close this ranking, with an average funding of USD 7.24, 10.96, and 12.73 mln, respectively. Given the significant heterogeneity across cybersecurity sectors in terms of the number of rounds, average funding, and total capital raised, our findings support the rejection of the null of H1a.

The total PMVs come at about one order of magnitude larger than the funding amount, which is typical of the private equity sector. The sectors' rank for PMVs is similar to the one of the total funding amount, albeit with some differences. Again, in this case, artificial intelligence and machine learning represent the most prominent sectors, with a total of PMVs recorded throughout the period at the trillion USD order of magnitude. However, this statistic should be interpreted with caution in this case as it may report the sequences of several valuations provided over the company lifecycle. In contrast to the funding descriptive statistics, the private cloud sector ranks first, with an average firm valuation of USD 447.22 mln, followed by the intrusion detection sector (USD 303.36 mln), QR codes (USD 300.61), and fraud detection (USD 251.16 mln). For the average company in these sectors, one USD invested translates into up to USD 10 in PMVs. These tenfold returns between two funding rounds are typically in the order of magnitude advertised by private equity funds. However, these substantial returns should be taken cautiously because of the large standard deviations associated with these categories' valuations. For instance, for the private cloud and fraud detection sectors, it is USD 1,406 mln and USD 1,338 mln, respectively. Second, our data almost surely come with a selection bias and an over-representation of \quotes{unicorns}. Nonetheless, we do not see any fundamental reason for this selection bias not to be uniformly distributed across categories. Thus, the ranking we observe should represent each sector's relative performance. Overall, we find that the valuations of cybersecurity firms vastly differ depending on the sector they belong to. These findings strongly support the rejection of H1b. Finally, we also find a substantial standard deviation associated with valuations in all sectors. Despite the observations of valuations do not match our methodology to compute returns from funding rounds to exits, it further justifies using a log model in the subsequent analysis. This is because returns computed from these valuations will also likely be affected.

\begin{table}[H]
\centering
\resizebox{0.8\textwidth}{!}{%
\begin{tabular}{lccccccccc}
& &\multicolumn{4}{c}{Funding amount} & \multicolumn{4}{c}{PMVs}\\ \cmidrule(lr){3-6} \cmidrule(lr){7-10}
Sector & \#Rounds & Avg. & Median  & Total & SD & Avg. & Median  & Total & SD \\ \cmidrule(lr){1-1} \cmidrule(lr){2-2} \cmidrule(lr){3-3} \cmidrule(lr){4-4} \cmidrule(lr){5-5} \cmidrule(lr){6-6} \cmidrule(lr){7-7} \cmidrule(lr){8-8} \cmidrule(lr){9-9} \cmidrule(lr){10-10} 

Artificial intelligence & $6,339$ & $19.60$ & $4.10$ & $124,256.80$ & $79.64$ & $151.64$ & $21.21$ & $961,248.10$ & $686.76$ \\ 
Biometrics              & $130$ & $10.96$ & $3.80$ & $1,425.26$ & $28.50$ & $67.26$ & $20.00$ & $8,743.24$ & $200.00$ \\ 
Blockchain              & $1,392$ & $19.47$ & $3.10$ & $27,104.71$ & $71.19$ & $199.31$ & $18.70$ & $277,443.60$ & $891.12$ \\ 
Cloud security          & $502$ & $25.23$ & $10.00$ & $12,664.88$ & $47.17$ & $206.52$ & $48.81$ & $103,671.70$ & $564.79$ \\ 
Cyber security          & $1,759$ & $22.71$ & $8.42$ & $39,946.45$ & $49.50$ & $202.40$ & $39.31$ & $356,017.50$ & $655.71$ \\ 
E-signature             & $41$ & $16.97$ & $2.23$ & $695.79$ & $48.93$ & $132.17$ & $11.23$ & $5,418.92$ & $476.42$ \\ 
Facial recognition      & $60$ & $31.48$ & $3.85$ & $1,888.50$ & $114.33$ & $201.32$ & $17.43$ & $12,078.95$ & $646.72$ \\ 
Fraud detection         & $223$ & $26.72$ & $8.00$ & $5,959.54$ & $74.31$ & $251.16$ & $39.04$ & $56,009.71$ & $1,388.45$ \\  
Internet of Things      & $1,768$ & $12.73$ & $3.00$ & $22,507.51$ & $42.35$ & $87.04$ & $16.75$ & $153,880.50$ & $287.91$ \\ 
Intrusion detection     & $29$ & $24.50$ & $11.00$ & $710.52$ & $50.72$ & $303.36$ & $51.80$ & $8,797.36$ & $730.08$ \\ 
Machine learning        & $3,974$ & $16.93$ & $4.50$ & $67,288.87$ & $48.17$ & $134.42$ & $22.50$ & $534,165.90$ & $474.34$ \\ 
Network security        & $1,034$ & $21.71$ & $10.00$ & $22,448.87$ & $43.03$ & $170.78$ & $46.97$ & $176,586.40$ & $442.26$ \\ 
Penetration testing     & $28$ & $14.54$ & $5.67$ & $407.06$ & $28.20$ & $89.31$ & $33.21$ & $2,500.68$ & $191.99$ \\ 
Privacy                 & $238$ & $18.03$ & $5.35$ & $4,291.04$ & $48.61$ & $158.94$ & $27.43$ & $37,827.69$ & $613.25$ \\ 
Private cloud           & $63$ & $45.46$ & $11.00$ & $2,863.77$ & $108.43$ & $447.22$ & $56.84$ & $28,174.61$ & $1,406.49$ \\ 
QR codes                & $17$ & $37.76$ & $2.00$ & $641.88$ & $86.38$ & $300.61$ & $10.71$ & $5,110.34$ & $695.72$ \\ 
Quantum computing       & $78$ & $21.41$ & $9.74$ & $1,669.95$ & $53.79$ & $142.90$ & $44.06$ & $11,146.37$ & $381.99$ \\ 
Security                & $3,551$ & $19.02$ & $7.51$ & $67,543.82$ & $41.89$  & $153.43$ & $35.33$ & $544,825.10$ & $501.43$ \\ 
Spam filtering          & $8$ & $7.24$ & $6.25$ & $57.90$ & $6.41$ & $32.39$ & $26.88$ & $259.14$ & $28.36$ \\ 
Total                   & $21,234$ & $19.04$ & $5.00$ & $404,373.10$ & $60.36$ & $154.65$ & $26.19$ & $3,283,906.00$ & $609.43$\\ 
\end{tabular}}
\caption{Descriptive statistics of funding amount PMVs for cybersecurity sectors}
\footnotesize{This table reports the number of funding rounds and average, median, total, and standard deviation estimates for the funding amount and PMVs regarding the 19 cybersecurity sectors. The values are in USD mln, and the study period is 2010--2022.}
\label{tab_descr}
\end{table}


Next, we report in Table \ref{IPO_Investors} descriptive statistics regarding the average time to IPO from the first funding round observation available.\footnote{We thank an anonymous referee of The 22nd Workshop on the Economics of Information Security (WEIS) for this suggestion.} With over 3,400 months (almost 10 years), the E-signature is, on average, the sector taking the longest to reach an IPO, followed by the fraud detection and cloud cybersecurity sectors (almost seven years each). Conversely, the blockchain sector only needs less than three and a half years to reach IPO from the first recorded round, followed by the spam filtering and the facial recognition sectors. We also present the average number of investors per round, for which the cross-sectional variations are much more limited. For instance, whereas the sector with the highest average number of investors is five (penetration testing), that with the lowest number is about three (spam filtering).


\begin{table}[H] 
\centering 
\resizebox{0.8\textwidth}{!}{%
\begin{tabular}{lcccccc}
 & \multicolumn{4}{c}{Time to IPO (days)} & \multicolumn{2}{c}{Number of investors}\\ \cmidrule(lr){2-5} \cmidrule(lr){6-7}
 Sectors & Avg. & SD & Max & Min & Avg. & SD\\ \cmidrule(lr){1-1} \cmidrule(lr){2-2} \cmidrule(lr){3-3} \cmidrule(lr){4-4} \cmidrule(lr){5-5} \cmidrule(lr){6-6} \cmidrule(lr){7-7}
Artificial Intelligence & $1,819$ & $1,019$ & $5,509$ & $52$ & $3.78$ & $4.08$ \\ 
Biometrics & $2,018$ & $1,069$ & $4,011$ & $47$ & $3.76$ & $3.64$ \\ 
Blockchain & $1,203$ & $895$ & $4,050$ & $65$ & $3.62$ & $3.54$ \\ 
Cloud Security & $2,570$ & $1,509$ & $6,173$ & $260$ & $4.22$ & $3.90$ \\ 
Cyber Security & $2,462$ & $1,616$ & $8,169$ & $119$ & $3.79$ & $3.94$ \\ 
E-Signature & $3,413$ & $1,775$ & $5,065$ & $595$ & $3.84$ & $3.86$ \\ 
Facial Recognition & $1,626$ & $1,168$ & $3,744$ & $196$ & $3.75$ & $3.34$ \\ 
Fraud Detection & $2,583$ & $1,643$ & $5,268$ & $166$ & $3.64$ & $3.27$ \\ 
Internet of Things & $1,927$ & $1,268$ & $6,454$ & $7$ & $3.63$ & $3.65$ \\ 
Intrusion Detection & $1,986$ & $1,265$ & $2,881$ & $1,092$ & $3.09$ & $2.91$ \\ 
Machine Learning & $1,907$ & $1,010$ & $5,573$ & $82$ & $3.74$ & $3.76$ \\ 
Network Security & $2,364$ & $1,394$ & $7,344$ & $119$ & $3.66$ & $3.74$ \\ 
Penetration Testing & NA & NA & NA & NA & $5.03$ & $4.31$ \\ 
Privacy & $1,642$ & $1,643$ & $5,858$ & $31$ & $3.58$ & $3.45$ \\ 
Private Cloud & $2,096$ & $1,070$ & $3,608$ & $206$ & $3.69$ & $3.32$ \\ 
QR Codes & NA & NA & NA & NA & $5.15$ & $3.47$ \\ 
Quantum Computing & $1,962$ & $397$ & $2,252$ & $1,528$ & $4.08$ & $3.55$ \\ 
Security & $2,413$ & $1,511$ & $8,169$ & $47$ & $3.77$ & $3.92$ \\ 
Spam Filtering & $1,574$ & $649$ & $1,949$ & $825$ & $2.91$ & $2.88$ \\ 
\end{tabular}}

\caption{Descriptive statistics of time to IPO and number of investors for cybersecurity sectors}
\footnotesize{This table reports descriptive statistics about the time to IPO (means, standard deviations, maximum and minimum time). We discard the sectors with less than 10 IPO observations. It also reports the means and standard deviations of the number of investors. The study period is 2010--2022.}
\label{IPO_Investors} 
\end{table}

\subsection{Statistical differences in funding and PMVs across categories}

Table \ref{tab_ttest_funding} reports the t-statistics of average tests for unequal variances, comparing average funding in each sector. The private cloud sector's average funding is statically superior to 12 sectors at the 10\% level and to six sectors at the 5\% level. Based on this criterion, the QR code sector ranks second, although it is not statistically superior to any other sector at the usual significance levels. In contrast, the spam filtering sector ranks last, inferior to 10 sectors at the 1\% level, followed by the biometrics sector (inferior to seven at the 1\% level).

\begin{table}[H] 
\centering 
 \resizebox{\textwidth}{!}{%
\begin{tabular}{@{\extracolsep{5pt}} lccccccccccccccccccc} 
Sectors/Sectors & AI & BI & BL & CS & CS & E-S & FR & FD & IoT & ID & ML & NS & PT & PR & PC & QR & QC & S & SF \\ \cmidrule(lr){1-1}\cmidrule(lr){2-2}\cmidrule(lr){3-3}\cmidrule(lr){4-4}\cmidrule(lr){5-5}\cmidrule(lr){6-6}\cmidrule(lr){7-7}\cmidrule(lr){8-8}\cmidrule(lr){9-9}\cmidrule(lr){10-10}\cmidrule(lr){11-11}\cmidrule(lr){12-12}\cmidrule(lr){13-13}\cmidrule(lr){14-14}\cmidrule(lr){15-15}\cmidrule(lr){16-16}\cmidrule(lr){17-17}\cmidrule(lr){18-18}\cmidrule(lr){19-19}\cmidrule(lr){20-20} 

Artificial Intelligence &  & $3.21$ & $0.06$ & $-2.41$ & $-2.01$ & $0.34$ & $-0.80$ & $-1.40$ & $4.84$ & $-0.52$ & $2.12$ & $-1.26$ & $0.93$ & $0.48$ & $-1.89$ & $-0.87$ & $-0.29$ & $0.48$ & $4.99$ \\ 
Biometrics & $-3.21$ &  & $-2.71$ & $-4.36$ & $-4.25$ & $-0.75$ & $-1.37$ & $-2.83$ & $-0.66$ & $-1.39$ & $-2.28$ & $-3.79$ & $-0.61$ & $-1.76$ & $-2.48$ & $-1.27$ & $-1.59$ & $-3.10$ & $1.10$ \\ 
Blockchain & $-0.06$ & $2.71$ &  & $-2.03$ & $-1.44$ & $0.32$ & $-0.81$ & $-1.36$ & $3.12$ & $-0.52$ & $1.24$ & $-0.96$ & $0.87$ & $0.39$ & $-1.88$ & $-0.87$ & $-0.30$ & $0.22$ & $4.13$ \\ 
Cloud Security & $2.41$ & $4.36$ & $2.03$ &  & $1.04$ & $1.04$ & $-0.42$ & $-0.28$ & $5.35$ & $0.08$ & $3.70$ & $1.41$ & $1.87$ & $1.90$ & $-1.46$ & $-0.60$ & $0.59$ & $2.80$ & $5.82$ \\ 
Cyber Security & $2.01$ & $4.25$ & $1.44$ & $-1.04$ &  & $0.74$ & $-0.59$ & $-0.79$ & $6.43$ & $-0.19$ & $4.11$ & $0.56$ & $1.50$ & $1.39$ & $-1.66$ & $-0.72$ & $0.21$ & $2.68$ & $6.06$ \\ 
E-Signature & $-0.34$ & $0.75$ & $-0.32$ & $-1.04$ & $-0.74$ &  & $-0.87$ & $-1.07$ & $0.55$ & $-0.62$ & $0.005$ & $-0.61$ & $0.26$ & $-0.13$ & $-1.82$ & $-0.93$ & $-0.45$ & $-0.27$ & $1.22$ \\ 
Facial Recognition & $0.80$ & $1.37$ & $0.81$ & $0.42$ & $0.59$ & $0.87$ &  & $0.31$ & $1.27$ & $0.40$ & $0.98$ & $0.66$ & $1.08$ & $0.89$ & $-0.70$ & $-0.25$ & $0.63$ & $0.84$ & $1.62$ \\ 
Fraud Detection & $1.40$ & $2.83$ & $1.36$ & $0.28$ & $0.79$ & $1.07$ & $-0.31$ &  & $2.76$ & $0.21$ & $1.95$ & $0.97$ & $1.67$ & $1.48$ & $-1.29$ & $-0.51$ & $0.68$ & $1.53$ & $3.56$ \\ 
Internet of Things & $-4.84$ & $0.66$ & $-3.12$ & $-5.35$ & $-6.43$ & $-0.55$ & $-1.27$ & $-2.76$ &  & $-1.24$ & $-3.32$ & $-5.36$ & $-0.33$ & $-1.60$ & $-2.39$ & $-1.19$ & $-1.41$ & $-5.12$ & $2.22$ \\ 
Intrusion Detection & $0.52$ & $1.39$ & $0.52$ & $-0.08$ & $0.19$ & $0.62$ & $-0.40$ & $-0.21$ & $1.24$ &  & $0.80$ & $0.29$ & $0.92$ & $0.65$ & $-1.26$ & $-0.58$ & $0.28$ & $0.58$ & $1.78$ \\ 
Machine Learning & $-2.12$ & $2.28$ & $-1.24$ & $-3.70$ & $-4.11$ & $-0.005$ & $-0.98$ & $-1.95$ & $3.32$ & $-0.80$ &  & $-3.10$ & $0.44$ & $-0.34$ & $-2.08$ & $-0.99$ & $-0.73$ & $-2.01$ & $4.06$ \\ 
Network Security & $1.26$ & $3.79$ & $0.96$ & $-1.41$ & $-0.56$ & $0.61$ & $-0.66$ & $-0.97$ & $5.36$ & $-0.29$ & $3.10$ &  & $1.31$ & $1.08$ & $-1.73$ & $-0.76$ & $0.05$ & $1.78$ & $5.50$ \\ 
Penetration Testing & $-0.93$ & $0.61$ & $-0.87$ & $-1.87$ & $-1.50$ & $-0.26$ & $-1.08$ & $-1.67$ & $0.33$ & $-0.92$ & $-0.44$ & $-1.31$ &  & $-0.56$ & $-2.11$ & $-1.07$ & $-0.85$ & $-0.83$ & $1.26$ \\ 
Privacy & $-0.48$ & $1.76$ & $-0.39$ & $-1.90$ & $-1.39$ & $0.13$ & $-0.89$ & $-1.48$ & $1.60$ & $-0.65$ & $0.34$ & $-1.08$ & $0.56$ &  & $-1.96$ & $-0.93$ & $-0.49$ & $-0.31$ & $2.78$ \\ 
Private Cloud & $1.89$ & $2.48$ & $1.88$ & $1.46$ & $1.66$ & $1.82$ & $0.70$ & $1.29$ & $2.39$ & $1.26$ & $2.08$ & $1.73$ & $2.11$ & $1.96$ &  & $0.31$ & $1.61$ & $1.93$ & $2.76$ \\ 
QR Codes & $0.87$ & $1.27$ & $0.87$ & $0.60$ & $0.72$ & $0.93$ & $0.25$ & $0.51$ & $1.19$ & $0.58$ & $0.99$ & $0.76$ & $1.07$ & $0.93$ & $-0.31$ &  & $0.75$ & $0.89$ & $1.45$ \\ 
Quantum Computing & $0.29$ & $1.59$ & $0.30$ & $-0.59$ & $-0.21$ & $0.45$ & $-0.63$ & $-0.68$ & $1.41$ & $-0.28$ & $0.73$ & $-0.05$ & $0.85$ & $0.49$ & $-1.61$ & $-0.75$ &  & $0.39$ & $2.18$ \\ 
Security & $-0.48$ & $3.10$ & $-0.22$ & $-2.80$ & $-2.68$ & $0.27$ & $-0.84$ & $-1.53$ & $5.12$ & $-0.58$ & $2.01$ & $-1.78$ & $0.83$ & $0.31$ & $-1.93$ & $-0.89$ & $-0.39$ &  & $4.97$ \\ 
Spam Filtering & $-4.99$ & $-1.10$ & $-4.13$ & $-5.82$ & $-6.06$ & $-1.22$ & $-1.62$ & $-3.56$ & $-2.22$ & $-1.78$ & $-4.06$ & $-5.50$ & $-1.26$ & $-2.78$ & $-2.76$ & $-1.45$ & $-2.18$ & $-4.97$ &  \\ 
\end{tabular}}
 \caption{T-tests of unequal variances of funding amount for each category}
 \footnotesize{This table reports t-statistics for pairwise mean t-tests of funding amount per sector. The t-statistics correspond to tests for unequal (Welch's t-tests) variance. The values are computed by comparing the row-wise to the column-wise entries}
  \label{tab_ttest_funding} 
\end{table}

Table \ref{tab_ttest_pmv} repeats the above analysis for pairwise comparison of PMVs per sector. Based on this criterion, we find the intrusion detection sector to strictly dominate other sectors, with average valuations being statistically superior at the 10\% level for 12 out of 18 comparisons and statistically superior at the 1\% level for seven comparisons. The private cloud sector ranks second and strictly dominates all technologies with the same significance (except for intrusion detection). Instead, the lowest ranking sectors are the E-signature, statistically inferior at the 10\% level to nine technologies and at the 1\% level to eight. Firms follow this low rank in the QR-codes sector. Finally, the pairwise statistical tests of average funding and PMVs support the rejection of H1a and H1b, respectively.

\begin{table}[H] 
\centering 
 \resizebox{\textwidth}{!}{%
\begin{tabular}{@{\extracolsep{5pt}} lccccccccccccccccccc} 
 Sectors/Sectors& AI & BI & BL & CS & CS & E-S & FR & FD & IoT & ID & ML & NS & PT & PR & PC & QR & QC & S & SF \\ \cmidrule(lr){1-1}\cmidrule(lr){2-2}\cmidrule(lr){3-3}\cmidrule(lr){4-4}\cmidrule(lr){5-5}\cmidrule(lr){6-6}\cmidrule(lr){7-7}\cmidrule(lr){8-8}\cmidrule(lr){9-9}\cmidrule(lr){10-10}\cmidrule(lr){11-11}\cmidrule(lr){12-12}\cmidrule(lr){13-13}\cmidrule(lr){14-14}\cmidrule(lr){15-15}\cmidrule(lr){16-16}\cmidrule(lr){17-17}\cmidrule(lr){18-18}\cmidrule(lr){19-19}
Artificial Intelligence &  & $1.63$ & $2.32$ & $-9.56$ & $-12.89$ & $1.29$ & $0.42$ & $-6.46$ & $7.31$ & $-3.27$ & $-0.56$ & $-13.99$ & $-0.45$ & $-3.42$ & $-3.80$ & $0.66$ & $-2.96$ & $-12.13$ & $0.03$ \\ 
Biometrics & $-1.43$ &  & $-0.69$ & $-6.27$ & $-5.75$ & $0.48$ & $-0.46$ & $-5.40$ & $0.81$ & $-3.67$ & $-1.77$ & $-6.57$ & $-1.11$ & $-3.42$ & $-4.15$ & $0.36$ & $-3.38$ & $-4.70$ & $-0.68$ \\ 
Blockchain & $-2.36$ & $0.61$ &  & $-9.76$ & $-11.33$ & $0.86$ & $-0.11$ & $-7.06$ & $3.39$ & $-3.65$ & $-2.58$ & $-12.54$ & $-0.87$ & $-4.27$ & $-4.23$ & $0.50$ & $-3.55$ & $-9.87$ & $-0.40$ \\ 
Cloud security & $9.08$ & $5.98$ & $9.39$ &  & $1.87$ & $3.72$ & $3.45$ & $0.26$ & $12.76$ & $-0.72$ & $9.13$ & $0.34$ & $2.03$ & $3.16$ & $-0.82$ & $1.63$ & $1.07$ & $4.04$ & $2.53$ \\ 
Cyber security & $12.42$ & $5.39$ & $11.44$ & $-1.88$ &  & $3.27$ & $2.90$ & $-1.08$ & $16.31$ & $-1.27$ & $11.79$ & $-2.11$ & $1.54$ & $2.20$ & $-1.47$ & $1.42$ & $0.27$ & $3.39$ & $2.05$ \\ 
E-signature & $-1.36$ & $-0.53$ & $-0.89$ & $-4.10$ & $-3.67$ &  & $-0.75$ & $-3.49$ & $-0.14$ & $-3.24$ & $-1.36$ & $-3.69$ & $-1.24$ & $-2.36$ & $-3.47$ & $0.16$ & $-2.67$ & $-2.73$ & $-0.90$ \\ 
Facial recognition & $-0.43$ & $0.49$ & $0.11$ & $-3.66$ & $-3.13$ & $0.75$ &  & $-3.16$ & $1.01$ & $-2.83$ & $-0.51$ & $-3.43$ & $-0.62$ & $-1.77$ & $-3.09$ & $0.51$ & $-2.16$ & $-2.23$ & $-0.25$ \\ 
Fraud detection & $5.89$ & $5.34$ & $6.47$ & $-0.26$ & $1.05$ & $3.88$ & $3.37$ &  & $9.11$ & $-0.81$ & $6.22$ & $-0.04$ & $1.85$ & $2.45$ & $-0.92$ & $1.57$ & $0.83$ & $2.58$ & $2.34$ \\ 
Internet of Things & $-7.15$ & $-0.75$ & $-3.41$ & $-12.61$ & $-16.30$ & $0.15$ & $-1.07$ & $-8.70$ &  & $-4.39$ & $-7.33$ & $-17.18$ & $-1.60$ & $-6.39$ & $-5.10$ & $0.22$ & $-4.75$ & $-15.66$ & $-1.14$ \\ 
Intrusion detection & $2.92$ & $3.73$ & $3.20$ & $0.69$ & $1.21$ & $3.14$ & $2.71$ & $0.80$ & $4.08$ &  & $3.20$ & $0.84$ & $2.00$ & $1.97$ & $0.01$ & $1.79$ & $1.24$ & $1.83$ & $2.36$ \\ 
Machine learning & $0.55$ & $1.60$ & $2.65$ & $-8.94$ & $-11.64$ & $1.48$ & $0.53$ & $-5.86$ & $7.31$ & $-2.95$ &  & $-13.01$ & $-0.39$ & $-3.19$ & $-3.72$ & $0.69$ & $-2.84$ & $-10.57$ & $0.10$ \\ 
Network security & $12.27$ & $6.79$ & $12.17$ & $-0.36$ & $2.05$ & $4.60$ & $4.09$ & $0.04$ & $16.54$ & $-0.89$ & $11.92$ &  & $1.97$ & $3.31$ & $-0.96$ & $1.59$ & $0.96$ & $5.45$ & $2.48$ \\ 
Penetration testing & $0.40$ & $1.11$ & $0.74$ & $-1.89$ & $-1.43$ & $1.19$ & $0.59$ & $-1.80$ & $1.45$ & $-2.00$ & $0.35$ & $-2.05$ &  & $-0.70$ & $-2.15$ & $0.78$ & $-1.18$ & $-0.99$ & $0.34$ \\ 
Privacy & $3.00$ & $3.42$ & $3.81$ & $-3.07$ & $-2.07$ & $2.71$ & $1.95$ & $-2.45$ & $5.91$ & $-2.02$ & $2.90$ & $-3.40$ & $0.71$ &  & $-2.26$ & $1.11$ & $-0.91$ & $-0.71$ & $1.18$ \\ 
Private cloud & $4.27$ & $4.53$ & $4.62$ & $0.95$ & $1.74$ & $3.43$ & $3.08$ & $1.03$ & $5.90$ & $-0.01$ & $4.31$ & $1.25$ & $1.95$ & $2.62$ &  & $1.82$ & $1.37$ & $2.12$ & $2.54$ \\ 
QR codes & $-1.17$ & $-0.61$ & $-0.86$ & $-2.92$ & $-2.67$ & $-0.19$ & $-0.69$ & $-2.76$ & $-0.40$ & $-2.11$ & $-1.26$ & $-3.29$ & $-0.91$ & $-2.02$ & $-2.35$ &  & $-1.33$ & $-1.21$ & $-0.61$ \\ 
Quantum computing & $2.65$ & $3.41$ & $3.15$ & $-1.03$ & $-0.26$ & $2.81$ & $2.20$ & $-0.82$ & $4.44$ & $-1.24$ & $2.64$ & $-1.02$ & $1.16$ & $0.92$ & $-1.41$ & $2.00$ &  & $0.62$ & $1.61$ \\ 
Security & $11.79$ & $4.53$ & $10.41$ & $-4.17$ & $-3.43$ & $3.18$ & $2.50$ & $-2.58$ & $15.99$ & $-1.80$ & $10.53$ & $-5.22$ & $0.96$ & $0.69$ & $-2.62$ & $2.37$ & $-0.61$ &  & $1.50$ \\ 
Spam filtering & $-0.01$ & $0.39$ & $0.18$ & $-1.27$ & $-1.01$ & $0.55$ & $0.14$ & $-1.26$ & $0.55$ & $-1.66$ & $-0.05$ & $-1.38$ & $-0.25$ & $-0.65$ & $-1.35$ & $0.44$ & $-0.95$ & $-0.76$ &  \\ 
\end{tabular}}
\caption{Cybersecurity sectors pairwise comparison of the average PMVs}
\footnotesize{This table reports t-statistics for pairwise mean t-tests of PMVs per sector. The t-statistics correspond to tests for unequal (Welch's t-tests) variance. The values are computed by comparing the row-wise to the column-wise entries, and the period is 2010--2022}
\label{tab_ttest_pmv} 
\end{table}

\subsection{Trends in cybersecurity venture capital}

Table \ref{tab_percent_change} reports the annualized average and standard deviations for the quarterly percentage changes in funding amount and PMVs.\footnote{In Appendix, Figure \ref{fig_funding_share}, we also depict the quarterly evolutions of funding amount and PMVs for the 19 selected sectors.} The most significant annualized increase in terms of funding amount is for the fraud detection sector, with over 283\%, followed by cloud security (263.59\%) and privacy (190.45\%). The top rank in funding change is consistent, with over 586\% for fraud detection, 251.59\% for cloud security, and 192.42\% for privacy. We understand that these astonishing figures mostly come from the fact that our database records coverage improves over time and that some of the sectors depicted were virtually inexistent at the beginning of the sample period. This is further confirmed by the massive standard deviations associated with these annualized average percent changes. We also observe that the lowest annualized average percent changes are associated with the most significant sector. Indeed, artificial intelligence, security, and machine learning are subject to a 32.19\%, 50.26\%, and 68.11\% average percent changes in funding and 52.07\%, 48.87\%, and 69.42\% in terms of PMVs.

\begin{table}[!htbp] \centering 
 \resizebox{0.8\textwidth}{!}{%
\begin{tabular}{@{\extracolsep{5pt}} lcccc}
& \multicolumn{2}{c}{Funding amount} & \multicolumn{2}{c}{PMVs}\\\cmidrule(lr){2-3} \cmidrule(lr){4-5} 
 & Avg. percent change & SD & Avg. percent change & SD \\\cmidrule(lr){2-2} \cmidrule(lr){3-3} \cmidrule(lr){4-4} \cmidrule(lr){5-5} 
Artificial intelligence & $32.19$ & $92.15$ & $52.07$ & $107.82$ \\ 
Blockchain & $107.77$ & $148.65$ & $145.84$ & $202.51$ \\ 
Cloud security & $263.59$ & $378.79$ & $251.59$ & $370.77$ \\ 
Cyber security & $50.24$ & $108.90$ & $87.58$ & $146.13$ \\  
Fraud detection & $283.38$ & $516.20$ & $586.02$ & $970.53$ \\ 
Internet of Things & $74.93$ & $157.41$ & $137.15$ & $237.00$ \\ 
Machine learning & $68.11$ & $105.91$ & $69.42$ & $121.38$ \\ 
Network security & $84.70$ & $154.42$ & $66.72$ & $131.87$ \\ 
Privacy & $190.45$ & $321.21$ & $192.42$ & $361.88$ \\ 
Security & $50.26$ & $111.96$ & $48.87$ & $112.29$ \\ 
\end{tabular}}

 \caption{Quarterly percent changes statistics for funding amount and PMVs}
 \footnotesize{This table reports, for each sector, the annualized quarterly percent changes for funding and PMVs and their annualized standard deviations. We discard the sectors with more than 20 missing quarterly observations. The period is 2010-2022.}
\label{tab_percent_change}
\end{table}

\subsection{Cybersecurity ventures locations}

\begin{figure}[!htbp]
        \centering
	\includegraphics[width=0.7\linewidth]{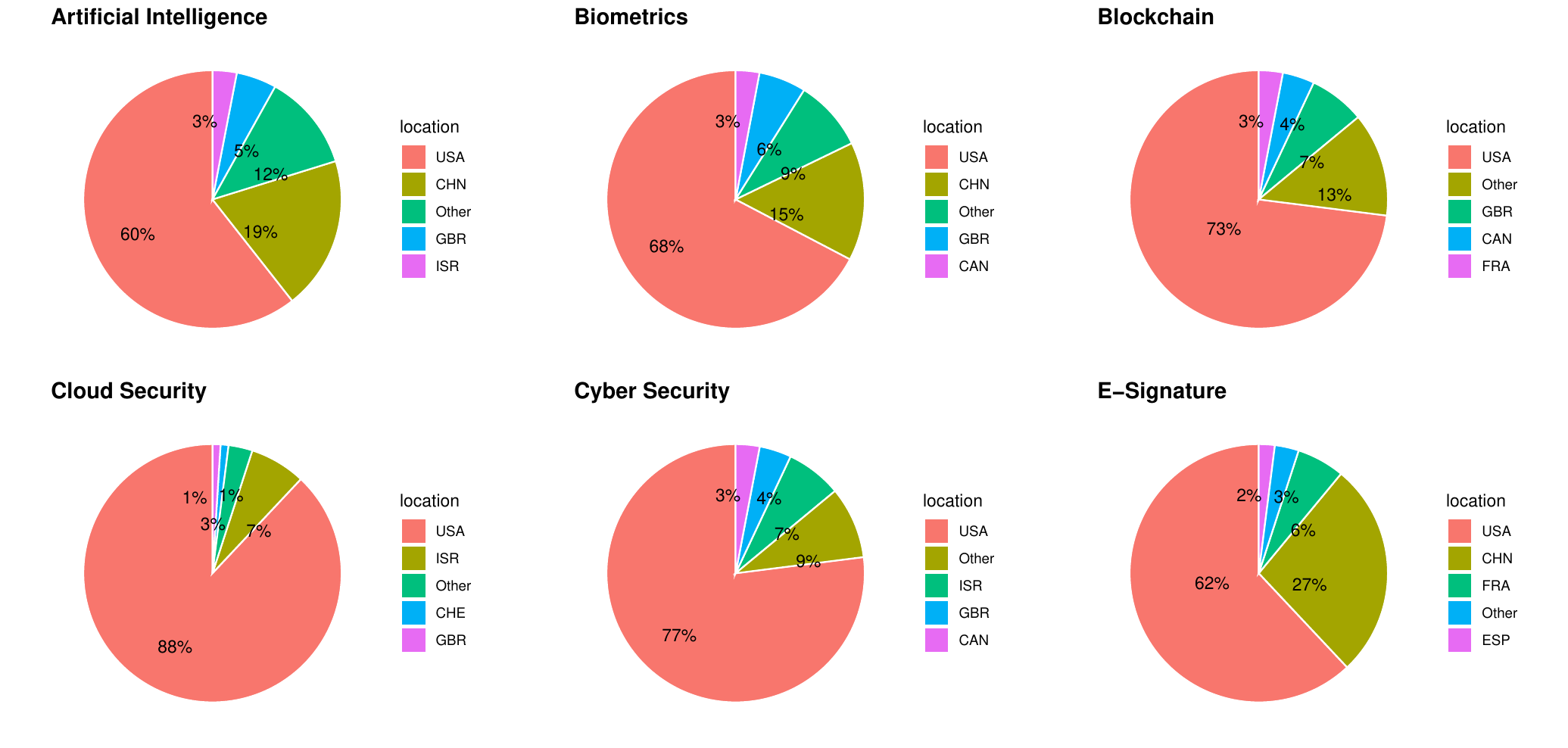}
	\includegraphics[width=0.7\linewidth]{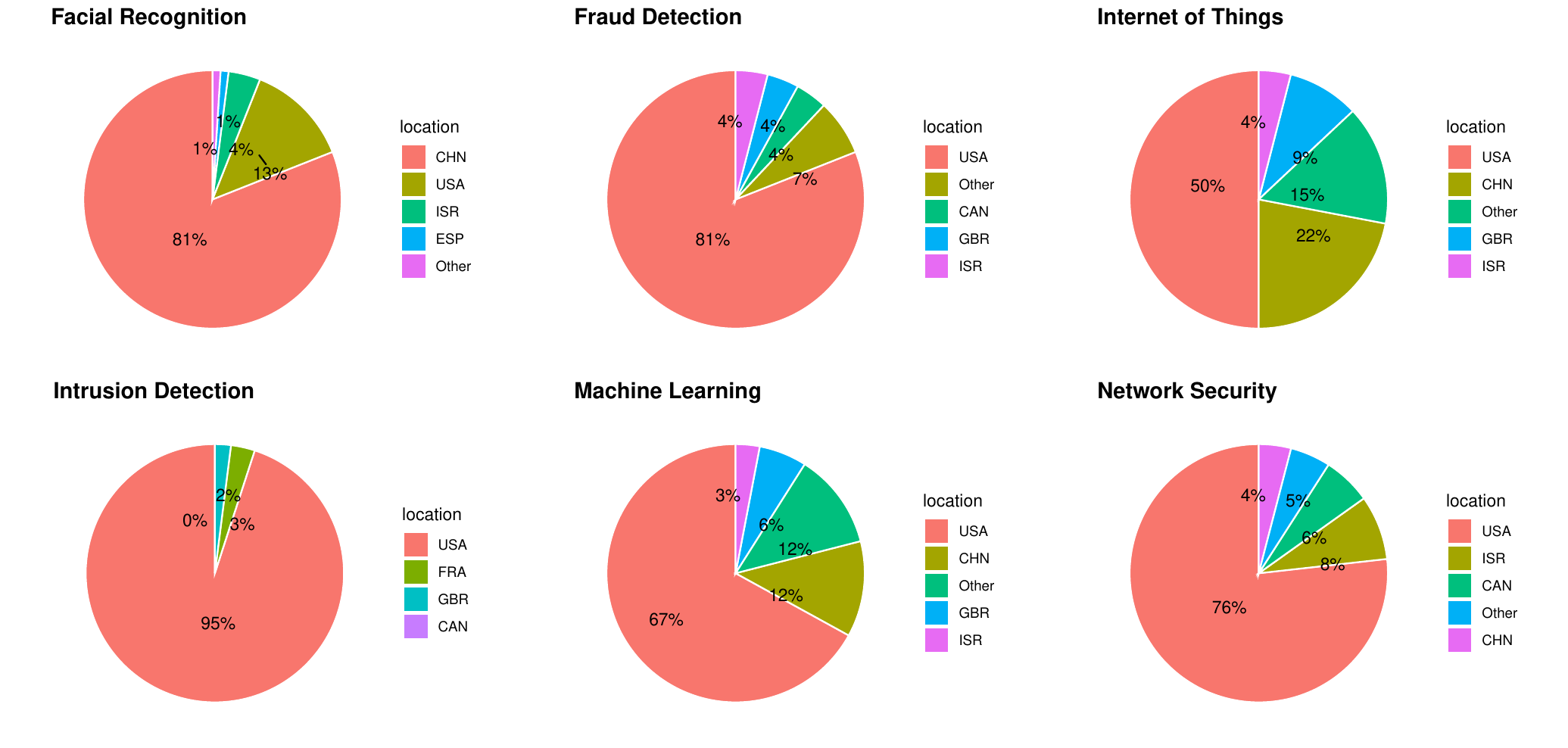}
	\includegraphics[width=0.7\linewidth]{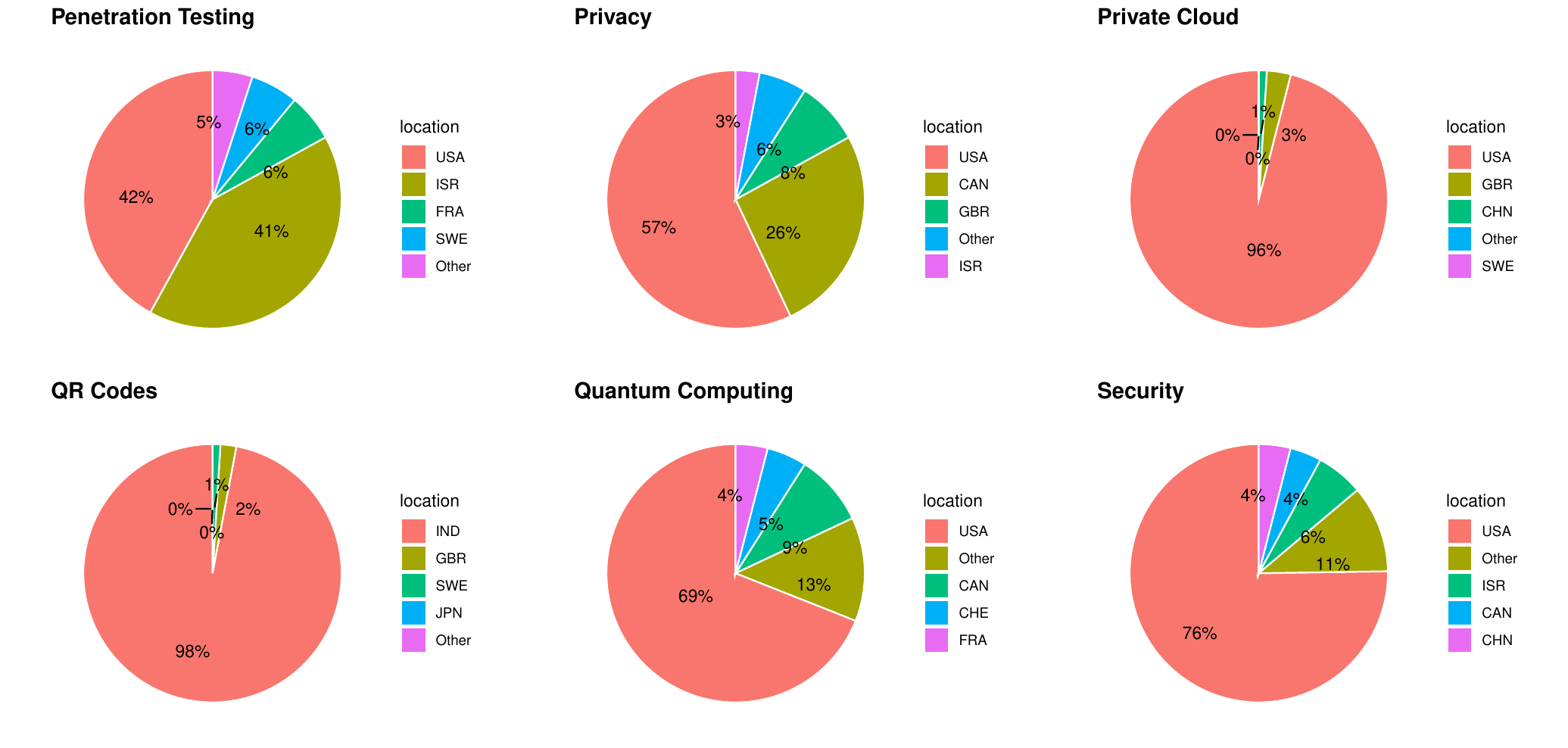}
    \caption{Location repartition of funded firms' capital raised in cybersecurity sectors}
    \footnotesize{This figure reports the share of funding amount per country where the firms raising capital have their headquarters. We report the top four locations ranked by funding amount and aggregate all the others in an \quotes{Other} category. We do not report the spam filtering sector since we find in the dataset that it is 100\% US-based. The period of observations is 2010--2022.}
    \label{fig_funding_countries}
\end{figure}

In Figure \ref{fig_funding_countries}, we display the share of investment raised by firms of each sector, depending on their top five locations. Unsurprisingly, the US accounts for the vast majority of this investment, with the exceptions of facial recognition, where China alone has more than 80\% of the firms targeted by private equity investments.\footnote{This is perhaps not surprising given the massive adoption of facial recognition for its security and the implementation of the social credit system in China.} The other sector vastly dominated by a non-US country is the QR codes sector, where India's firms alone represent 98\% of the amount invested.\footnote{We also try to make sense of these surprising figures, and from anecdotal evidence, we find support for massive adoption of QR codes, primarily for payment systems. See, \textit{e.g.}, \href{https://www.bloomberg.com/press-releases/2022-06-03/digital-payments-in-india-projected-to-reach-10-trillion-by-2026-phonepe-pulse-and-bcg-release-report-on-digital-payments-l3yea50y}{https://www.bloomberg.com/press-releases/2022-06-03/digital-payments-in-india}} A third technology for which the US dominance is not clear is the penetration testing sector. Indeed, whereas US firms in this sector receive 42\% of investment, it is closely followed by Israel with 41\% of the funding. By the same token, whereas US privacy firms are leaders in capital raised, with 57\%, Canadian firms capture a significant part of this investment with 26\%. Unsurprisingly, China often reaches a second, considerable market share, with 27\% for E-signature, 22\% for IoT, 12\% for machine learning, 19\% for artificial intelligence, and 15\% for biometrics firms.

\subsection{Financial performance of cybersecurity ventures}

\subsubsection{Log model regressions}

Using Cochrane's (2005) approach, we compute the returns to exit, accounting for capital dilution across firms' lifecycle. Next, we set up a three months grid to average returns on each sector. We match the quarterly returns on the S\&P 500 and the risk-free rate observations at the quarterly frequency and estimate the log model of Eq \ref{eq_2}. As in Table \ref{tab_percent_change}, we must discard technologies for which there are not enough observations to estimate the model or, more importantly, that are not sufficiently spread over 2010--2022. We are left with 12 sectors.
We report the results of the parameters' estimations, $\gamma$, $\delta$, and $\sigma$, as well as the implied parameters $\alpha$ and $\beta$ for the arithmetic form. We find $\alpha$, the return in excess of the market risk premium, to be positive for all technologies. Even though we cannot compute the standard errors for these implied parameters, the corresponding parameter in the log model $\gamma$ is significant at the 1\% level for nine cybersecurity sectors, except privacy and private cloud (5\% level significance) and blockchain (non statistically significant). In terms of size, this intercept translates into between 4.28\% (network security) and 5.16\% (blockchain) of annualized risk-adjusted returns. The $\delta$ parameter is not statistically significant at the usual levels, except for the privacy sector, at the 1\% level. The implied parameter $\beta$ ranges between 0.51 for the biometrics sector and 5.51 for the privacy sector. This parameter measures the covariance of the selected sector with the index. It evaluates the systematic risk amount to which the sector is exposed in the standard Capital Asset Pricing Model framework. The larger this coefficient, the higher the expected returns, everything being equal. This parameter also provides information about the sector's cyclicality with the broad economy. The larger (smaller) the coefficient, the more pro-(contra-) cyclical the sector is. This is particularly important for cybersecurity sectors since, as other sectors that mitigate crises, they tend to be more contra-cyclical. One anecdotal evidence is the performance of the pharmaceutical sector in the COVID crisis. As the broad economy collapsed, pharma companies experienced, on average, high returns, with COVID treatment prospects. With this consideration, the artificial intelligence sector, with $\beta$ over 1.62, is mainly pro-cyclical, which is not surprising given the non-exclusivity of this sector for cybersecurity. Instead, firms for which the activity in the cybersecurity sector is more explicit, those tagged as cybersecurity, network security, fraud detection, and private cloud, have $\beta$ well below one. Instead, making sense of the privacy sector's outstandingly significant $\beta$ is hard.

\begin{table}[!htbp] \centering 
 \resizebox{0.7\textwidth}{!}{%
\begin{tabular}{@{\extracolsep{5pt}} lccccccc} 

Sector & $\gamma$ & se($\gamma$) & $\delta$ & se($\delta$) & $\sigma$ (\%) &  $\alpha$ & $\beta$\\ \cmidrule(lr){1-1}\cmidrule(lr){2-2}\cmidrule(lr){3-3}\cmidrule(lr){4-4}\cmidrule(lr){5-5}\cmidrule(lr){6-6}\cmidrule(lr){7-7}\cmidrule(lr){8-8}
Artificial intelligence      & $0.13^{***}$ & $0.03$ & $0.48$ & $0.47$ & $31.24$ & $1.14$ & $1.62$ \\ 
Biometrics      & $0.07^{***}$ & $0.02$ & $-0.68$ & $0.63$ & $5.24$ & $1.08$ & $0.51$ \\ 
Blockchain  & $0.26$ & $0.20$ & $-0.26$ & $8.95$ & $91.80$ & $1.29$ & $0.77$ \\ 
Cloud security  & $0.11^{***}$ & $0.01$ & $0.25$ & $0.19$ & $11.24$ & $1.12$ & $1.28$ \\ 
Cyber security   & $0.08^{***}$ & $0.01$ & $-0.32$ & $0.18$ & $13.71$ & $1.08$ & $0.73$ \\ 
E-signature  & $0.08^{**}$ & $0.04$ & $0.49$ & $1.02$ & $6.73$ & $1.09$ & $1.63$ \\
Fraud detection & $0.13^{***}$ & $0.05$ & $-0.56$ & $0.73$ & $17.33$ & $1.14$ & $0.57$ \\ 
Internet of Things & $0.07^{***}$ & $0.02$ & $-0.35$ & $0.35$ & $20.31$ & $1.08$ & $0.71$ \\ 
Machine learning  & $0.13^{***}$ & $0.02$ & $0.15$ & $0.30$ & $19.13$ & $1.13$ & $1.17$ \\ 
Network security & $0.07^{***}$ & $0.01$ & $-0.20$ & $0.21$ & $14.79$ & $1.07$ & $0.82$ \\ 
Privacy & $0.12^{**}$ & $0.05$ & $1.71^{***}$ & $0.65$ & $17.74$ & $1.13$ & $5.51$\\ 
Private cloud & $0.12^{**}$ & $0.06$ & $-0.12$ & $0.74$ & $15.38$ & $1.13$ & $0.88$\\
All sectors & $0.07^{***}$ & $0.02$ & $0.48^{***}$ & $0.03$ & $14.18$ & $1.07$ & $1.62$ \\ 

\end{tabular}}
 \caption{Estimates for the log model regressions and the implied values for $\alpha$ and $\beta$.}
\footnotesize{This table reports the estimations of the parameters from regression of excess quarterly log-returns on the market risk premium, approximated by the excess returns of the S\&P 500 on the risk-free rate (see Eq. \ref{eq_2}), for 12 cybersecurity sectors. We report the exact standard errors for the $\gamma$ and $\delta$ parameters and the total volatility in percentage. We also report the $\alpha$ and $\beta$ parameters, equivalent of $\gamma$ and $\delta$ in the model in arithmetic form. $^{***}$, $^{**}$, $^{*}$ indicates significativity at the 1\%, 5\%, and 10\% levels, respectively. The period of observations is 2010--2022.}
 
\label{tap_regression_parameters} 
\end{table} 

\subsubsection{Implied expected returns}

Table \ref{tab_implied_exp_returns} presents the expectation and standard errors for log returns computed with Eqs. \ref{eq_3}, \ref{eq_4}, and the corresponding one for the arithmetic returns computed with Eqs. \ref{eq_5}, \ref{eq_6}. The blockchain sector reaches annual expected arithmetic (log) returns of 177.27\% (105.42\%), which is in line with the underlying performance of cryptocurrencies over the 2010-2022 period. The second sector in this ranking is artificial intelligence, with annualized returns of 67.25\%, in line with the results of Cochrane (2005) for the info. sector (79\%). Other sectors in the high range include machine learning (58.5\% p.a.), private cloud (57.88\% p.a.), and cloud security (50.58\% p.a.). Therefore, we reject the null of H3, as we find support for substantial variations in risk-adjusted returns, systematic risk, and expected returns. We thus find strong support to reject the null of H3, with substantial heterogeneity across sectors, in terms of systematic risks ($\beta$ from 0.71 to 5.51) and expected returns (from 9.72\% p.a. to 177.27\% p.a.). In contrast, we find that risk-adjusted returns (annualized $\alpha$ from 4.28\% to 5.16\%) lie in a tight range. Together with their relatively small sizes, as opposed to non-risk-adjusted returns, this also points to correct pricing in the cybersecurity sector. Overall, the total risk, risk-adjusted returns, expected returns, and systematic risks differ vastly across sectors. These results in financial performance parameters strongly support the rejection of the null of H2. Finally, the fact that all risk-adjusted returns are positive points to the overall attractiveness of the cybersecurity industry.

\begin{table}[!htbp] \centering 
\resizebox{0.7\textwidth}{!}{%
\begin{tabular}{@{\extracolsep{5pt}} lcccc}
Sector & E{[}ln R{]} &  se(E{[}ln R{]}) & E{[}R{]} & se(E{[}R{]}) \\ \cmidrule(lr){1-1} \cmidrule(lr){2-2} \cmidrule(lr){3-3} \cmidrule(lr){4-4} \cmidrule(lr){5-5}
Artificial intelligence & $55.55$ & $14.62$ & $67.25$ & $16.85$ \\ 
Biometrics & $31.30$ & $4.72$ & $23.22$ & $2.55$ \\ 
Blockchain & $105.42$ & $42.15$ & $177.27$ & $64.14$ \\ 
Cloud security & $47.21$ & $5.37$ & $50.58$ & $5.81$ \\ 
Cyber security & $33.09$ & $6.57$ & $36.40$ & $6.87$ \\ 
E-signature & $36.10$ & $4.25$ & $48.54$ & $3.46$ \\ 
Fraud detection & $54.99$ & $8.63$ & $46.34$ & $8.89$ \\ 
Internet of Things & $31.87$ & $9.55$ & $37.77$ & $10.23$ \\ 
Machine learning & $52.57$ & $8.83$ & $58.50$ & $10.09$ \\ 
Network security & $29.41$ & $6.89$ & $31.81$ & $7.34$ \\ 
Privacy & $51.89$ & $13.07$ & $9.72$ & $8.35$ \\ 
Private cloud & $50.81$ & $7.09$ & $57.88$ & $8.09$ \\
All sectors & $34.17$ & $8.48$ & $37.43$ & $6.94$ \\ 

\end{tabular}}
\caption{Implied estimates for $\mathbb{E}[\ln R]$, and $\mathbb{E}[R]$}
\footnotesize{This table reports the implied estimates for the expected value and standard errors computed from the log and arithmetic models with quarterly returns in 12 cybersecurity sectors. We annualize the values and display them in percentages. The study period is 2010--2022.}
  \label{tab_implied_exp_returns} 
\end{table} 

Compared to previous findings of the private equity literature, the cybersecurity industry stands out. Cochrane (2005) finds an average annualized arithmetic return of 59\% for all industries combined, ranging from 42\% for the health sector to 111\% for the retail sector~\cite{Cochrane2005}. Our average arithmetic returns for the cybersecurity industry are lower, with only 37\%. However, compared individually, some cybersecurity sectors, such as artificial intelligence and blockchain, reach the top of this range or completely dominate it. Cochrane (2005) additionally finds that the broad level of systematic risk, corresponding to the $\beta$ parameter, is 1.9 for the whole private equity market and ranges across industries between -0.1 (retail) and 1.7 (info.). We find similar values, with 1.62 on an aggregate, and the range of systematic risk across cybersecurity sectors falls in the same range, except for the privacy sector. Ewens (2009) finds values slightly closer to ours, with a beta of 2.4 and an annualized alpha of 27\%~\cite{Ewens2009}. Finally, with a methodology closer to ours, Peng (2004) built a venture capital index spanning 1987--1999. He finds an average return of 55.18\% per year and an index $\beta$ with the S\&P500 of 2.4~\cite{Peng2001W}. Thus, would the private equity markets behave similarly across the period, our results would support rejecting the null hypothesis H3.

\section{Conclusion} \label{conclusion}


Using financial data, we examine the financial performance of private firms operating in cybersecurity sectors. We focus on 19 cybersecurity-related sectors identified in Crunchbase and analyze their funding rounds, firm valuations, and exits (IPOs or acquisitions) from 2010 to 2022. While Crunchbase provides exhaustive data on funding rounds, many valuations and IPOs share price observations are missing. Therefore, we use a machine learning approach to estimate these values. Using Cochrane's (2005) approach, we compute returns accounting for capital dilutions since several funding rounds before exits are common in the private equity sector. Overall, the artificial intelligence, security, and machine learning sectors raised over USD 60 bln to USD 120 bln in funding. Other sectors rank better based on other criteria, such as average and median, particularly the private cloud sector. We confirm these results with a mean t-test analysis, finding that the intrusion detection sector dominates other sectors in post-money valuations, followed by the private cloud sector. We also examine trends by analyzing percent changes over the sample period and finding that the fraud detection sector experienced the most significant increase in funding and valuations, followed by cloud security and privacy. We sort capital raised and valuations by geographical locations, with US firms dominating most sectors, except for facial recognition (dominated by China) and QR Codes (dominated by India). However, China also captures a significant market share in several sectors, such as E-signature, IoT, and biometrics. Finally, we estimate parameters and compute implied parameters for 12 cybersecurity sectors using a log return process and calculate returns to exit. The $\alpha$ parameter, return in excess of the market risk premium, is positive for all sectors, while the $\beta$ parameter, indicating the systematic risk and cyclicality of the sectors, ranges between 0.51 and 5.51. We find sectors such as artificial intelligence to be pro-cyclical, while firms in more explicit cybersecurity sectors have $\beta$ well below one, indicative of contra-cyclicality. We compute the expected arithmetic and log returns from the model parameters. We find blockchain to have the highest expected annual arithmetic and log returns, followed by artificial intelligence, machine learning, private cloud, and cloud security. Privacy and biometrics rank the lowest in terms of expected returns. Overall our results contribute to the view that cybersecurity sectors are highly heterogeneous regarding financial performance, regardless of the criterion used to benchmark them. Moreover, depending on the criterion, some sectors may rank well in some performance metrics and poorly in others.


One of the caveats of our research comes from the limited amount of observations. Although we are confident of our ML approach to infer missing PMVs from funding round observations, we cannot discard the fact that Crunchbase may not record all the actual funding rounds. Since larger funding rounds are more likely to be reported than smaller ones, we acknowledge the presence of a selection bias in our sample. Another limitation comes from the generation of unbalanced sample sizes across technologies. It is an issue for statistical inferences when the number of observations is too small and prevents us from estimating parameters in some cybersecurity sectors. We are also aware that we analyze the financial performance of these sectors in a period starting when some of them were virtually inexistent and which reached a significant economic size in both private and public equity. This explains, for instance, the tremendous annualized average increase of each cybersecurity sector in terms of capital raised and PMVs.


To address the aforementioned limitations, we plan to improve this research in several directions. In particular, we aim to render our dataset more exhaustive. Although Crunchbase aggregates several data sources, it is far from complete, in particular for the sectors studied, given that they are among the most recent ones. The replication of this study in about five years would translate into at least a 50\% increase in our current sample size and helps to make inference in the long run. 
Next, using natural language processing models, we would like to map the Crunchbase taxonomy to existing ones, such as ENISA. In addition, this mapping could be penalized to maximize the balancing of our observations across sectors if needed. Last, using Crunchbase's additional features, or traditional asset pricing risk factors, we aim to explain the cross-sectional difference in sectors' financial performance. These explanatory variables could also be chosen for technology monitoring, such as the number of patents, scientific articles, or GitHub repositories per cybersecurity sector.



\newpage
\bibliographystyle{styles/jfe}
\bibliography{references.bib}

\begin{thebibliography}{72}
\expandafter\ifx\csname natexlab\endcsname\relax\def\natexlab#1{#1}\fi

\bibitem[{Agrafiotis et~al.(2018)Agrafiotis, Nurse, Goldsmith, Creese, and
  Upton}]{AgrafiotisNurseGoldsmithCreeseUpton2018}
Agrafiotis, I., Nurse, J. R.~C., Goldsmith, M., Creese, S., Upton, D., 2018. A
  taxonomy of cyber-harms: Defining the impacts of cyber-attacks and
  understanding how they propagate. Journal of Cybersecurity 4, 1--15.

\bibitem[{Alexy et~al.(2012)Alexy, Block, Sandner, and
  Ter~Wal}]{AlexyBlockSandnerTerwal2012}
Alexy, O.~T., Block, J.~H., Sandner, P., Ter~Wal, A. L.~J., 2012. Social
  capital of venture capitalists and start-up funding. Small Business Economics
  39, 835--851.

\bibitem[{Amir et~al.(2018)Amir, Levi, and Livne}]{AmirLeviLivne2018}
Amir, E., Levi, S., Livne, T., 2018. Do firms underreport information on
  cyber-attacks? {E}vidence from capital markets. Review of Accounting Studies
  23, 1177--1206.

\bibitem[{Anderson et~al.(2013)Anderson, Barton, Böhme, Clayton,
  Eeten~\(van\), Levi, Moore, and
  Savage}]{AndersonBartonBöhmeClaytonEetenLeviMooreSavage2013}
Anderson, R., Barton, C., Böhme, R., Clayton, R., Eeten~\(van\), M. J.~G.,
  Levi, M., Moore, T., Savage, S., 2013. Measuring the cost of cybercrime.
  Workshop on the Economics of Information Security 11, 265--300.

\bibitem[{Anderson et~al.(2019)Anderson, Barton, Böhme, Clayton, Ganan,
  Grasso, Levi, Moore, and
  Vasek}]{AndersonBartonBöhmeClaytonGananGrassoLeviMooreVasek2019}
Anderson, R., Barton, C., Böhme, R., Clayton, R., Ganan, C., Grasso, T., Levi,
  M., Moore, T., Vasek, M., 2019. Measuring the changing cost of cybercrime.
  Workshop on the Economics of Information Security 18, 1--32.

\bibitem[{Anderson and Moore(2006)}]{AndersonMoore2006}
Anderson, R., Moore, T., 2006. The economics of information security. Science
  314, 610--613.

\bibitem[{Ang et~al.(2018)Ang, Chen, Goetzmann, and
  Phalippou}]{AngChenGoetzmannPhalippou2018}
Ang, A., Chen, B., Goetzmann, W.~N., Phalippou, L., 2018. Estimating private
  equity returns from limited partner cash flows. Journal of Finance 73,
  1751--1783.

\bibitem[{Axelson and Martinovic(2015)}]{AxelsonMartinovic2015}
Axelson, U., Martinovic, M., 2015. European venture capital: Myths and facts.
  Available at
  \url{https://personal.lse.ac.uk/axelson/ulf_files/EuroVC_MythsFacts\%20v17.pdf}\EatDot
  .

\bibitem[{Baryshnikov(2012)}]{Baryshnikov2012}
Baryshnikov, Y., 2012. {IT} security investment and {G}ordon-{L}oeb's {1\/e}
  rule. Workshop on Economics and Information Security 11, 1--6.

\bibitem[{Besten~\(den\)(2021)}]{Besten2021}
Besten~\(den\), M.~L., 2021. Crunchbase research: Monitoring entrepreneurship
  research in the age of big data. Available at
  \url{http://dx.doi.org/10.2139/ssrn.3724395}\EatDot .

\bibitem[{Bodin et~al.(2008)Bodin, Gordon, and Loeb}]{BodinGordonLoeb2008}
Bodin, L.~D., Gordon, L.~A., Loeb, M.~P., 2008. Information security and risk
  management. Communications of the ACM 51, 64--48.

\bibitem[{Boehmer et~al.(1991)Boehmer, Musumecci, and
  Poulsen}]{BoehmerMusumecciPoulsen1991}
Boehmer, E., Musumecci, J., Poulsen, A.~B., 1991. Event-study methodology under
  conditions of event-induced variance. Journal of Financial Economics 30,
  253--272.

\bibitem[{Bouveret(2018)}]{Bouveret2018W}
Bouveret, A., 2018. Cyber risk for the financial sector: A framework for
  quantitative assessment. Available at
  \url{http://dx.doi.org/10.2139/ssrn.3203026}\EatDot .

\bibitem[{Campbell et~al.(1997)Campbell, Lo, and
  Mac{K}inlay}]{CampbellLoMacKinley1997}
Campbell, J.~Y., Lo, A.~W., Mac{K}inlay, A.~C., 1997. The Econometrics of
  Financial Markets. Princeton University Press, Princeton, New Jersey.

\bibitem[{Campbell et~al.(2003)Campbell, Gordon, Loeb, and
  Zhou}]{CampbellGordonLoebZhou2003}
Campbell, K., Gordon, L.~A., Loeb, M.~P., Zhou, L., 2003. The economic cost of
  publicly announced information security breaches: Empirical evidence from the
  stock market. Journal of Cybersecurity 11, 431--448.

\bibitem[{Cochrane(2005)}]{Cochrane2005}
Cochrane, J.~H., 2005. The risk and return of venture capital. Journal of
  Financial Economics 75, 3--52.

\bibitem[{Cumming and Dai(2010)}]{CummingDai2010}
Cumming, D., Dai, N., 2010. Local bias in venture capital investments. Journal
  of Empirical Finance 17, 362--380.

\bibitem[{Cumming and Dai(2011)}]{CummingDai2011}
Cumming, D., Dai, N., 2011. Fund size, limited attention and valuation of
  venture capital backed firms. Journal of Empirical Finance 18, 2--15.

\bibitem[{Cumming et~al.(2013)Cumming, Haß, and
  Schweizer}]{CummingHaßSchweizer2013}
Cumming, D., Haß, L.~H., Schweizer, D., 2013. Private equity benchmarks and
  portfolio optimization. Journal of Banking and Finance 37, 3515--3528.

\bibitem[{Dalle et~al.(2017)Dalle, Besten~\(den\), and
  Menon}]{DalleBestenMenon2017W}
Dalle, J.-M., Besten~\(den\), M.~L., Menon, C., 2017. Using {C}runchbase for
  economic and managerial research. Available at
  \url{https://doi.org/10.1787/18151965}\EatDot .

\bibitem[{Driessen et~al.(2012)Driessen, Lin, and
  Phalippou}]{DriessenLinPhalippou2012}
Driessen, J., Lin, T.-C., Phalippou, L., 2012. A new method to estimate risk
  and return of nontraded assets from cash flows: The case of private equity
  funds. Journal of Financial and Quantitative Analysis 47, 511--535.

\bibitem[{Engel and Keilbach(2007)}]{EngelKeilbach2007}
Engel, D., Keilbach, M., 2007. Firm-level implications of early stage venture
  capital investment - an empirical investigation. Journal of Empirical Finance
  14, 150--167.

\bibitem[{Ewens(2009)}]{Ewens2009}
Ewens, M., 2009. A new model of venture capital risk and return. Available at
  \url{http://dx.doi.org/10.2139/ssrn.1356322}\EatDot .

\bibitem[{Farrow and Szanton(2016)}]{FarrowSzanton2016}
Farrow, S., Szanton, J., 2016. Cybersecurity investment guidance: Extensions of
  the {G}ordon and {L}oeb {M}odel. Journal of Information Security 7, 15--28.

\bibitem[{Feurer et~al.(2021)Feurer, Eggensperger, Falkner, Lindauer, and
  Hutter}]{FeurerEtAl2021}
Feurer, M., Eggensperger, K., Falkner, S., Lindauer, M., Hutter, F., 2021.
  Auto-{Sklearn} 2.0: Hands-free {AutoML} via meta-learning. Available at
  \url{https://arxiv.org/abs/2007.04074}\EatDot .

\bibitem[{Fielder et~al.(2018)Fielder, König, Panaousis, Schauer, and
  Rass}]{FielderKonigPanaousisSchauerRass2018}
Fielder, A., König, S., Panaousis, E., Schauer, S., Rass, S., 2018. Risk
  assessment uncertainties in cybersecurity investments. Games 9, 1--14.

\bibitem[{Florackis et~al.(2023)Florackis, Louca, Michaely, and
  Weber}]{FlorackisLoucaMichaelyWeber2023}
Florackis, C., Louca, C., Michaely, R., Weber, M., 2023. Cybersecurity risk.
  Review of Financial Studies 36, 351--407.

\bibitem[{Franzoni et~al.(2012)Franzoni, Nowak, and
  Phalippou}]{FranzoniNowakPhalippou2012}
Franzoni, F., Nowak, E., Phalippou, L., 2012. Private equity performance and
  liquidity risk. Journal of Finance 67, 2341--2373.

\bibitem[{Gordon and Loeb(2002)}]{GordonLoeb2002}
Gordon, L.~A., Loeb, M.~P., 2002. The economics of information security
  investment. ACM Transactions on Information and System Security 5, 438--457.

\bibitem[{Gordon and Loeb(2006)}]{GordonLoeb2006}
Gordon, L.~A., Loeb, M.~P., 2006. Economic aspects of information security: An
  emerging field of research. Information Systems Frontiers 8, 335--337.

\bibitem[{Gordon et~al.(2015{\natexlab{a}})Gordon, Loeb, Lucyshyn, and
  Zhou}]{GordonLoebLucyshynZhou2015a}
Gordon, L.~A., Loeb, M.~P., Lucyshyn, W., Zhou, L., 2015{\natexlab{a}}.
  Externalities and the magnitude of cyber security underinvestment by private
  sector firms: A modification of the {G}ordon-{L}oeb {M}odel. Journal of
  Information Security 6, 24--30.

\bibitem[{Gordon et~al.(2015{\natexlab{b}})Gordon, Loeb, Lucyshyn, and
  Zhou}]{GordonLoebLucyshynZhou2015b}
Gordon, L.~A., Loeb, M.~P., Lucyshyn, W., Zhou, L., 2015{\natexlab{b}}. The
  impact of information sharing on cybersecurity underinvestment: A real
  options perspective. Journal of Accounting and Public Policy 34, 509--519.

\bibitem[{Gordon et~al.(2015{\natexlab{c}})Gordon, Loeb, Lucyshyn, and
  Zhou}]{GordonLoebLucyshynZhou2015c}
Gordon, L.~A., Loeb, M.~P., Lucyshyn, W., Zhou, L., 2015{\natexlab{c}}.
  Increasing cybersecurity investments in private sector firms. Journal of
  Cybersecurity 1, 3--17.

\bibitem[{Gordon et~al.(2003)Gordon, Loeb, and Sohail}]{GordonLoebSohail2003}
Gordon, L.~A., Loeb, M.~P., Sohail, T., 2003. A framework for using insurance
  for cyber-risk management. Communications of the ACM 46, 81--85.

\bibitem[{Gordon et~al.(2010)Gordon, Loeb, and Sohail}]{GordonLoebSohail2010}
Gordon, L.~A., Loeb, M.~P., Sohail, T., 2010. Market value of voluntary
  disclosures concerning information security. Management Information Systems
  Quarterly 34, 567--594.

\bibitem[{Gordon et~al.(2011)Gordon, Loeb, and Zhou}]{GordonLoebZhou2011}
Gordon, L.~A., Loeb, M.~P., Zhou, L., 2011. The impact of information security
  breaches: {H}as there been a downward shift in costs? Journal of Computer
  Security 19, 33--56.

\bibitem[{Gordon et~al.(2016)Gordon, Loeb, and Zhou}]{GordonLoebZhou2016}
Gordon, L.~A., Loeb, M.~P., Zhou, L., 2016. Investing in cybersecurity:
  Insights from the {G}ordon-{L}oeb {M}odel. Journal of Information Security 7,
  49--59.

\bibitem[{Gordon et~al.(2020)Gordon, Loeb, and Zhou}]{GordonLoebZhou2020}
Gordon, L.~A., Loeb, M.~P., Zhou, L., 2020. Integrating cost–benefit analysis
  into the {NIST} {C}ybersecurity {F}ramework via the {G}ordon–{L}oeb
  {M}odel. Journal of Cybersecurity 6, 1--8.

\bibitem[{Gordon et~al.(2018)Gordon, Loeb, and
  Zhou}]{GordonLoebLucyshynZhou2018}
Gordon, L.~A., Loeb, M. P.~Lucyshyn, W., Zhou, L., 2018. Empirical evidence on
  the determinants of cybersecurity investments in private sector firms.
  Journal of Information Security 9, 133--153.

\bibitem[{Gornall and Strebulaev(2020)}]{GornallStrebulaev2020}
Gornall, W., Strebulaev, I.~A., 2020. Squaring venture capital valuations with
  reality. Journal of Financial Economics 135, 120--143.

\bibitem[{Hausken(2006)}]{Hausken2006}
Hausken, K., 2006. Returns to information security investment: The effect of
  alternative information security breach functions on optimal investment and
  sensitivity to vulnerability. Information Systems Frontiers 8, 338--349.

\bibitem[{Hervé and Schwienbacher(2018)}]{HerveSchwienbacher2018}
Hervé, F., Schwienbacher, A., 2018. Round-number bias in investment: Evidence
  from equity crowdfunding. Finance 39 (1), 71--105.

\bibitem[{Hilary et~al.(2016)Hilary, Segal, and Zhang}]{HilarySegalZhang2016W}
Hilary, G., Segal, B., Zhang, M.~H., 2016. Cyber-risk disclosure: {W}ho cares?
  Available at \url{http://dx.doi.org/10.2139/ssrn.2852519}\EatDot .

\bibitem[{Hwang et~al.(2005)Hwang, Quigley, and
  Woodward}]{HwangQuigleyWoodward2005}
Hwang, M., Quigley, J.~M., Woodward, S.~E., 2005. An index for venture capital,
  1987--2003. Contributions to Economic Analysis and Policy 4, 1--43.

\bibitem[{Hwang et~al.(2022)Hwang, Shin, and Kim}]{HwangShinKim2022}
Hwang, S.-Y., Shin, D.-J., Kim, J.-J., 2022. Systematic review on
  identification and prediction of deep learning-based cyber security
  technology and convergence fields. Symmetry 14, 1--37.

\bibitem[{Johnson et~al.(2017)Johnson, Kang, and
  Lawson}]{JohnsonKangLawson2017}
Johnson, M., Kang, M.~J., Lawson, T., 2017. Stock price reaction to data
  breaches. Journal of Finance 16, 1--13.

\bibitem[{Kamiya et~al.(2021)Kamiya, Kang, Kim, Milidonis, and
  Stulz}]{KamiyaKangKimMilidonisStulz2021}
Kamiya, S., Kang, J.-K., Kim, J., Milidonis, A., Stulz, R.~M., 2021. Risk
  management, firm reputation, and the impact of successful cyberattacks on
  target firms. Journal of Financial Economics 139, 719--749.

\bibitem[{Kolari and Pynnonen(2010)}]{KolariPynnonen2010}
Kolari, J.~W., Pynnonen, S., 2010. Event study testing with cross-sectional
  correlation of abnormal returns. Review of Financial Studies 23, 3996--4025.

\bibitem[{Korteweg and Nagel(2016)}]{KortewegNagel2016}
Korteweg, A., Nagel, S., 2016. Risk‐adjusting the returns to venture capital.
  Journal of Finance 71, 1437--1470.

\bibitem[{Korteweg and Sorensen(2010)}]{KortewegSorensen2010}
Korteweg, A., Sorensen, M., 2010. Risk and return characteristics of venture
  capital-backed entrepreneurial companies. Review of Financial Studies 23,
  3738--3772.

\bibitem[{Laube and Böhme(2016)}]{LaubeBohme2016}
Laube, S., Böhme, R., 2016. The economics of mandatory security breach
  reporting to authorities. Journal of Cybersecurity 2, 29--41.

\bibitem[{Lelarge(2012)}]{Lelarge2012}
Lelarge, M., 2012. Coordination in network security games: A monotone
  comparative statics approach. IEEE Journal on Selected Areas in
  Communications 30, 2210--2219.

\bibitem[{Lending et~al.(2018)Lending, Minnick, and
  Schorno}]{LendingMinnickSchorno2018}
Lending, C., Minnick, K., Schorno, P.~J., 2018. Corporate governance, social
  responsibility, and data breaches. Financial Review 53, 413--455.

\bibitem[{McKenzie et~al.(2012)McKenzie, Satchell, and
  Wongwachara}]{McKenzieSatchellWongwachara2012}
McKenzie, M., Satchell, S., Wongwachara, W., 2012. Nonlinearity and smoothing
  in venture capital performance data. Journal of Empirical Finance 19,
  782--795.

\bibitem[{Mezzetti et~al.(2022)Mezzetti, Maréchal, Percia~David, Lacube,
  Gillard, Tsesmelis, Maillart, and
  Mermoud}]{MezzettiMarechalPercia-DavidLacubeGillardTsesmelisMaillartMermoud2022W}
Mezzetti, A., Maréchal, L., Percia~David, D., Lacube, W., Gillard, S.,
  Tsesmelis, M., Maillart, T., Mermoud, A., 2022. Techrank. Available at
  \url{https://arxiv.org/abs/2210.07824}\EatDot .

\bibitem[{Moore et~al.(2006)Moore, Dynes, and Chang}]{MooreDynesChang2016}
Moore, T., Dynes, S., Chang, F.~R., 2006. Identifying how firms manage
  cybersecurity investment. Workshop on the Economics of Information Security
  15, 1--27.

\bibitem[{Moskowitz and
  Vissing-Jørgensen(2002)}]{MoskowitzVissing-Jørgensen2002}
Moskowitz, T.~J., Vissing-Jørgensen, A., 2002. The returns to entrepreneurial
  investment: A private equity premium puzzle? American Economic Review 92,
  745--778.

\bibitem[{Peng(2001)}]{Peng2001W}
Peng, L., 2001. Building a venture capital index. Available at
  \url{http://dx.doi.org/10.2139/ssrn.281804}\EatDot .

\bibitem[{Poufinas and Vordonis(2018)}]{PoufinasVordonis2018}
Poufinas, T., Vordonis, N., 2018. Pricing the cost of cybercrime - {A}
  financial protection approach. iBusiness 10, 128--143.

\bibitem[{Romanosky(2016)}]{Romanosky2016}
Romanosky, S., 2016. Examining the costs and causes of cyber incidents. Journal
  of Cybersecurity 2, 121--135.

\bibitem[{Ruan(2017)}]{Ruan2017}
Ruan, K., 2017. Introducing cybernomics: A unifying economic framework for
  measuring cyber risk. Computers and Security 65, 77--89.

\bibitem[{Rue and Pfleeger(2009)}]{RuePfleeger2009}
Rue, R., Pfleeger, S.~L., 2009. Making the best use of cybersecurity economic
  models. IEEE Security and Privacy Magazine 7, 52--60.

\bibitem[{Schmidt(2006)}]{Schmidt2006}
Schmidt, D.~M., 2006. Private equity versus stocks. Journal of Alternative
  Investments 9 (1), 28--47.

\bibitem[{Shameli-Sendi et~al.(2016)Shameli-Sendi, Aghababaei-Barzegar, and
  Cheriet}]{Shameli-SendiAghababaei-BarzegarCheriet2016}
Shameli-Sendi, A., Aghababaei-Barzegar, R., Cheriet, M., 2016. Taxonomy of
  information security risk assessment ({ISRA}). Computers and Security 57,
  14--30.

\bibitem[{Spanos and Angelis(2016)}]{SpanosAngelis2016}
Spanos, G., Angelis, L., 2016. The impact of information security events to the
  stock market: A systematic literature review. Computers and Security 58,
  216--229.

\bibitem[{Tanaka et~al.(2005)Tanaka, Matsuura, and
  Sudoh}]{TanakaMatsuuraSudoh2005}
Tanaka, H., Matsuura, K., Sudoh, O., 2005. Vulnerability and information
  security investment: An empirical analysis of e-local government in {J}apan.
  Journal of Accounting and Public Policy 24, 37--59.

\bibitem[{Tosun(2021)}]{Tosun2021}
Tosun, O.~K., 2021. Cyber-attacks and stock market activity. International
  Review of Financial Analysis 76, 1--15.

\bibitem[{Wang et~al.(2008)Wang, Chaudhury, and Rao}]{WangChaudhuryRao2008}
Wang, J., Chaudhury, A., Rao, H.~R., 2008. Research note - {A} value-at-risk
  approach to information security investment. Information Systems Research 19,
  106--120.

\bibitem[{Wang(2019)}]{Wang2019}
Wang, S.~S., 2019. Integrated framework for information security investment and
  cyber insurance. Pacific-Basin Finance Journal 57, 1--12.

\bibitem[{Willemson(2006)}]{Willemson2006}
Willemson, J., 2006. On the {G}ordon and {L}oeb {M}odel for information
  security investment. Workshop on Economics and Information Security 5, 1--12.

\bibitem[{{‘Crunchbase, Inc.’}(2022)}]{Crunchbase2022}
{‘Crunchbase, Inc.’}, 2022. Historical company data. Crunchbase daily *.csv
  export, data retrieved on April 2022, from
  \url{https://data.crunchbase.com/docs/daily-csv-export}\EatDot.

\bibitem[{{‘National Institute of Standards and Technology,
  (NIST)‘}(2018)}]{NIST2018}
{‘National Institute of Standards and Technology, (NIST)‘}, 2018. Framework
  for improving critical infrastructure cybersecurity. Available at:
  \url{https://nvlpubs.nist.gov/nistpubs/cswpnist.cswp.04162018.pdf}\EatDot.

\end{thebibliography}

\newpage

\begin{appendices}
\section{Appendix}\label{appendix}
\begin{table}[H]
\resizebox{0.8\textwidth}{!}{%
\begin{tabular}{lll}
Value Stack Level 1 &
Value Stack Level 2 &
Value Stack Level 3 \\ \cmidrule(lr){1-1}  \cmidrule(lr){2-2}  \cmidrule(lr){3-3}
 & &\\
R\&D and education & Education & Cybersecurity academia/research \\
 & & Cybersecurity professional education \\
 & R\&D & Cyber threat and vulnerabilities research \\
 & & Cryptography research \\
 & & Software \& Hardware Research \& Development \\
 & & Cybersecurity standards development \\
 & & \\
Software &  Application security SW & Application Security Testing Software \\
 & & Vulnerability Assessment Software \\
 & & Web Application Firewalls Software \\
 & & Other Application Security Software \\
 & Cloud security SW & Cloud Access Security Brokers \\
 & & Cloud Security Posture Management \\
 & & Cloud Workload Protection Platforms \\
 & & Other Cloud Security Software \\
 & Data Security SW & Encryption Software \\
 & & Enterprise Data Loss Prevention Software \\
 & & Tokenization Software \\
 & & Other Data Security Software \\
 & & Software security Module \\
 & Identity and Access Management SW & Access Management Software \\
 & & Identity Governance and Administration Software \\
 & & Privileged Access Management Software \\
 & & User Authentication Software \\
 & Infrastructure Protection SW & Endpoint Protection Platform (Enterprise) Software \\
 & & Secure E-mail Gateway Software \\
 & & Secure Web Gateway Software \\
 & Operational software platforms & Security Information and Event Management (SIEM) Software \\
 & & Threat Intelligence Software \\
 & & Other Infrastructure Protection Software \\
 & Integrated Risk Management/GRC SW & Digital Risk Management (DRM) \\
 & & Vendor Risk Management (VRM) \\
 & & Business Continuity Management (BCM) \\
 & & Audit Management (AM) \\
 & & Corporate Compliance and Oversight (CCO) \\
 & & Enterprise Legal Management (ELM) \\
 & & Other Integrated Risk Management/GRC SW \\
 & & \\
Hardware & Network security equipment & Firewall Equipment, Intrusion Detection, and Prevention Systems\\
 & Hardware security & Trusted Platform Module \\
 & & Hardware security module \\
 & & Network security equipment \\
 & Biometric-based security equipment/systems & Hardware biometric security module \\
 & & Software biometric security module \\
 & & \\
 Distribution & Distribution & Software resale \\
 & & Hardware resale \\
 & & Managed Services resale \\
 & & \\
Advisory and consulting & Advisory and consulting &
Security and risk strategy, planning and management advice\\
 & & Security advisory and research \\
 & & Security testing and risk and threat assessment (Penetration Testing)\\
 & & Security Operations Centre (SOC) services\\
 & & Security Compliance and Audit (Compliance Management\\
 & & Digital forensics: post-event (incident/intrusion) analysis, Investigation\\
 & & Security project management, staff augmentation\\ 
 & & Other IT/cybersecurity consultancy services \\
 & & \\
Implementation services & Implementation design & Security design, engineering, and architecture development \\
 & Integration services & Implementation and integration, interoperability testing \\
 & Development & Implementation support (technical assistance/expert support services) \\
 & & \\
Managed Services & Managed response services & Managed Detection and Response (MDR) \\
 & Cybersecurity Device management & Security device management (including maintenance, patching\\
 & & Co-managed Services \\
 & Threats and Vulnerabilities &
Vulnerability Management \\
 & & Threat Detection Services\\ 
 & & Threat intelligence \\
 & Virtualized cybersecurity services & Cybersecurity as a service (CSaaS) \\
 & Security training & Security training services\\ 
 & Other managed services & Other Managed Services\\ 
 & & \\
Certification services & Product Cybersecurity Certification services & All services related to the assessment and implementation of assurance levels \\
 & Service/Process Certification services & All services related to the assessment and implementation of assurance levels\\ 
 & Professional Certification services & All services related to cybersecurity certification\\
& Accreditation services & All services related to the accreditation of cybersecurity\\
\end{tabular}}
\caption{ENISA taxonomy}
\footnotesize{This table reproduces the taxonomy of the European Union Agency for Cybersecurity (ENISA), precisely, from the ENISA cybersecurity market analysis framework (ECSMAF). See, \url{https://www.enisa.europa.eu/publications/enisa-cybersecurity-market-analysis-framework-ecsmaf}}
\label{tab:ENISA}
\end{table}
\newpage

\begin{table}[H]
\resizebox{\textwidth}{!}{%

\begin{tabular}{llll}
\textbf{Main category} & \textbf{Middle category} & \textbf{Small category} & \textbf{Hype cycle technology name}\\ \cmidrule(lr){1-1}  \cmidrule(lr){2-2}  \cmidrule(lr){3-3} \cmidrule(lr){4-4}
 & & Perimeter security &\\
 & & Secure connection &\\
 & \multirow{-3}{*}{Wired network security} & DDoS response & \multirow{-3}{*}{IDPS} \\
 & & Mobile communication network security &\\
 & \multirow{-2}{*}{Wireless network security} & Wireless local area network security & \multirow{-2}{*}{Mobile threat defense}\\
 & & Virtualization platform security &\\
 & & Cloud security service &\\
 \multirow{-8}{*}{ Network security} & \multirow{-3}{*}{Cloud security} & Software defined security & \multirow{-3}{*}{Cloud security assessments}\\
 & & & \\
 & & Web security &\\
 & & Email security &\\
 & \multirow{-3}{*}{Application security} & Database security & \multirow{-3}{*}{Secure web/gateways} \\
 & & Privacy protection &\\
 & & Data leakage prevention &\\
 & \multirow{-3}{*}{Data security} & Digital copyright infringement/right protection & \multirow{-3}{*}{Data loss prevention} \\
 & & E-money security &\\
 & & Blockchain security &\\
 & & Electronic transaction/abnormal behavior detection &\\
 & \multirow{-4}{*}{E-money, fintech security} & Prevention of transactions and fraud & \multirow{-4}{*}{The programmable economy} \\
 & & Digital evidence/collection and analysis &\\
 \multirow{-12}{*}{Data and application service security} & \multirow{-2}{*}{Digital forensics} & Anti-forensic response &\multirow{-2}{*}{E-Discovery software}\\
 & & &\\
 & & Biometric sensor &\\
 & & Biometric engine &\\
 & & Human recognition and search &\\
 & \multirow{-4}{*}{Human/bio-recognition} & Human/bio-recognition application & \multirow{-4}{*}{Biometric authentication methods}\\
 & & Camera and storage device &\\
 & & VMS/integrated control &\\
 & & Intelligent video surveillance &\\
 & \multirow{-4}{*}{CCTV surveillance/control} & CCTV infrastructure protection & \multirow{-4}{*}{Video/image}\\
 & & Interpersonal search machine &\\
 & & Luggage/luggage finder &\\
 & & Alarm monitoring &\\
\multirow{-12}{*}{Physical security} & \multirow{-4}{*}{Secure search and unmanned electronic guard} & Unmanned electronic security service & \multirow{-4}{*}{Analytics}\\
 & & &\\
 & & Password design &\\
 & & Cryptographic side channel analysis &\\
 & \multirow{-3}{*}{Cryptographic technique} & Password analysis & \multirow{-3}{*}{Database encryption}\\
 & & Universal authentication &\\
 & & ID management and access control &\\
 & \multirow{-3}{*}{Certification/authorization technique} & Bio certification & \multirow{-3}{*}{Externalized authorization management}\\
 & & SW vulnerability analysis &\\
 & \multirow{-2}{*}{Security vulnerability} & HW vulnerability analysis & \multirow{-2}{*}{Vulnerability assessment}\\
 & & Operating system security &\\
 & & Virtualization security &\\
 & \multirow{-3}{*}{System security} & System access control & \multirow{-3}{*}{Data loss prevention}\\
 & & Malware response &\\
 & \multirow{-2}{*}{Malware} & Ransomware response & \multirow{-2}{*}{DLP for mobile devices}\\
 & & Intelligent cyber threat analysis &\\
 & & Security information analysis and log management &\\
 \multirow{-16}{*}{System and password security} & \multirow{-3}{*}{Threat analysis and control} & Security control & \multirow{-3}{*}{SIEM}\\
 & & &\\
 & & Home city device security and control &\\
 & \multirow{-2}{*}{Home city security} & Home city data privacy & \multirow{-2}{*}{Smart city framework}\\
 & & Smart factory security &\\
 & & Infrastructure security &\\
 & \multirow{-3}{*}{Industrial control system security} & Smart energy security & \multirow{-3}{*}{Operational technology security}\\
 & & Communication security inside and outside the vehicle &\\
 & & Access control inside and outside the vehicle &\\
 & & Car intrusion detection &\\
 & \multirow{-4}{*}{Vehicle security} & Vehicle security vulnerability diagnosis & \multirow{-4}{*}{Mobile device integration into automobiles}\\
 & & Prevention of hacking of autonomous ships &\\
 & & Shipping port communication security &\\
 & & Marine infrastructure security control &\\
 & & Unmanned vehicle security &\\
 & \multirow{-5}{*}{Ship, ocean, and air security} & Aviation infrastructure security control &\multirow{-5}{*}{Autonomous vehicles}\\
 & & Healthcare device/sensor security &\\
 & \multirow{-2}{*}{Healthcare, medical security} & Medical data security and sharing & \multirow{-2}{*}{Real-time health system command center}\\
\multirow{-17}{*}{IoT security} & Other ICT security & Artificial intelligence and robot security & IoT securities\\
\end{tabular}}
\caption{Taxonomy of Hwang et al. (2022)}
\footnotesize{This table reproduces the taxonomy for cybersecurity sectors of Hwang et al. (2022) provided in their Table 4.\cite{HwangShinKim2022}}
\label{tab:Hwang}
\end{table}
\newpage
\clearpage

\begin{figure}[H]
    \centering
    \includegraphics[scale=0.5]{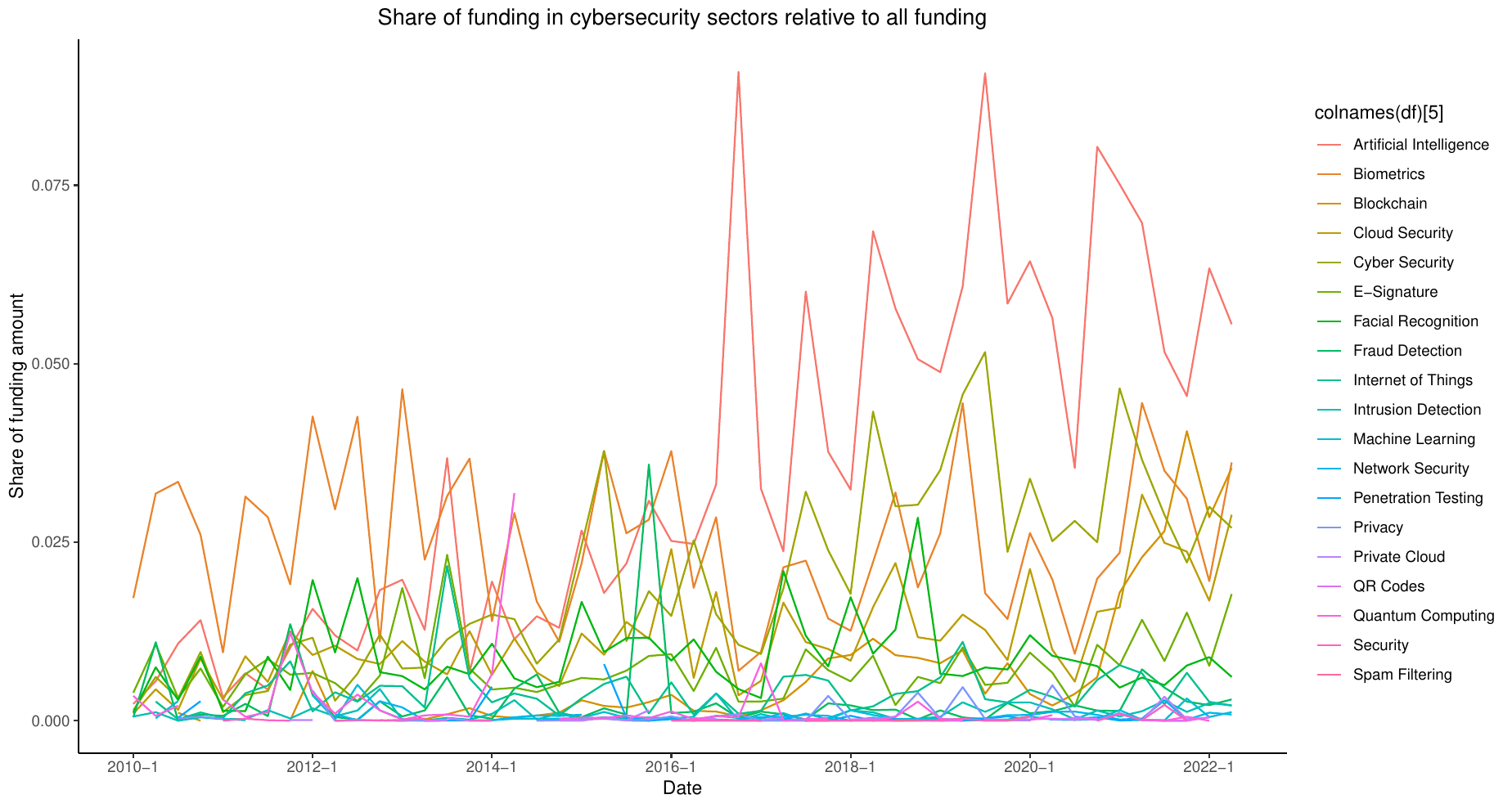}
    \caption{Share of capital raised in cybersecurity sectors over total capital raised in private equity}
    \footnotesize{This figure depicts the evolution of the capital raised in the 19 cybersecurity sectors, relative to the total capital raised in private equity. The frequency is quarterly, and the period is 2010--2022.} 
    \label{fig_funding_share}
\end{figure}

\begin{figure}[H]
    \centering
    \includegraphics[scale=0.5]{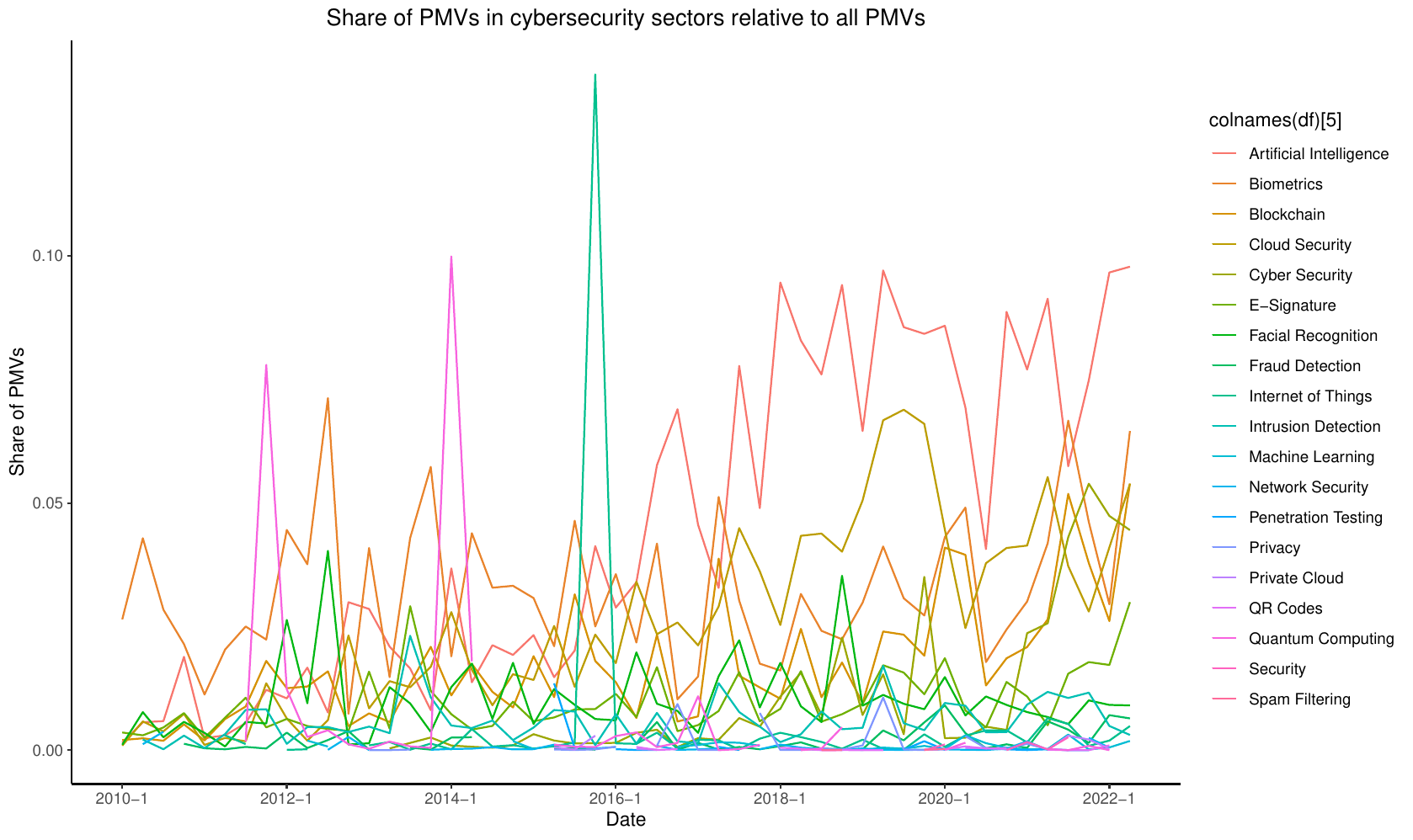}
    \caption{Share of PMVs in cybersecurity sectors over total capital raised in private equity}
    \footnotesize{This figure depicts the evolution of the PMVs of the 19 cybersecurity sectors, relative to the total PMVs in private equity. The frequency is quarterly, and the period is 2010--2022.} 
    \label{fig_pmv_share}
\end{figure}

\newpage

\end{appendices}

\end{document}